\documentclass[footinbib,aps,twocolumn,superscriptaddress,floatfix,prb]{revtex4-2}

\usepackage[utf8]{inputenc}
\usepackage{amssymb,amsmath,amsfonts,accents,amsthm}
\usepackage{mathtools}
\usepackage[colorlinks=true,citecolor=blue,linkcolor=blue,urlcolor=blue]{hyperref}
\usepackage{xcolor}
\usepackage{bbold}

\usepackage[german,english]{babel}

\makeatletter
\def\bbl@set@language#1{%
	\edef\languagename{%
		\ifnum\escapechar=\expandafter`\string#1\@empty
		\else\string#1\@empty\fi}%
	\@ifundefined{babel@language@alias@\languagename}{}{%
		\edef\languagename{\@nameuse{babel@language@alias@\languagename}}%
	}%
	\select@language{\languagename}%
	\expandafter\ifx\csname date\languagename\endcsname\relax\else
	\if@filesw
	\protected@write\@auxout{}{\string\select@language{\languagename}}%
	\bbl@for\bbl@tempa\BabelContentsFiles{%
		\addtocontents{\bbl@tempa}{\xstring\select@language{\languagename}}}%
	\bbl@usehooks{write}{}%
	\fi
	\fi}
\newcommand{\DeclareLanguageAlias}[2]{%
	\global\@namedef{babel@language@alias@#1}{#2}%
}
\makeatother

\DeclareLanguageAlias{en}{english}
\DeclareLanguageAlias{English}{english}
\DeclareLanguageAlias{Englisch}{english}
\DeclareLanguageAlias{EN}{english}
\DeclareLanguageAlias{en-US}{english}
\DeclareLanguageAlias{de}{german}

\usepackage{graphicx}
\usepackage[export]{adjustbox}

\newcommand{\PRLSec}[1]{\textit{#1}.---}

\usepackage{cleveref}

\newcommand{\sbb}{\overline{SS}}
\newcommand{\ssb}{S\sbar}
\newcommand{\sbs}{\sbar{}S}
\newcommand{\sbar}{\overline{S}}

\newcommand{\ISR}[2]{\widetilde{\ham}_{#1}(\lambda)}

\newcommand{\ham}{H}
\newcommand{\hC}{\ham}
\newcommand{\evecC}{\mathbf{Y}}
\newcommand{\evecA}{\mathbf{X}}
\newcommand{\evecAComponent}[1]{X_{#1}}
\newcommand{\evecCComponent}[1]{Y_{#1}}

\begin{document}
	\title{Hidden symmetries in acoustic wave systems}
	
	\author{M. Röntgen}
	\affiliation{%
		Zentrum für optische Quantentechnologien, Universität Hamburg, Luruper Chaussee 149, 22761 Hamburg, Germany
	}%
	
	\author{C. V. Morfonios}%
	\affiliation{%
		Zentrum für optische Quantentechnologien, Universität Hamburg, Luruper Chaussee 149, 22761 Hamburg, Germany
	}%
	
	\author{P. Schmelcher}
	\affiliation{%
		Zentrum für optische Quantentechnologien, Universität Hamburg, Luruper Chaussee 149, 22761 Hamburg, Germany
	}%
	\affiliation{%
		The Hamburg Centre for Ultrafast Imaging, Universität Hamburg, Luruper Chaussee 149, 22761 Hamburg, Germany
	}%
	
	\author{V. Pagneux}%
	\affiliation{%
		Laboratoire d’Acoustique de l’Université du Mans, Unite Mixte de Recherche 6613, Centre National de la Recherche Scientifique, Avenue O. Messiaen, F-72085 Le Mans Cedex 9, France
	}%

	\begin{abstract}
Latent symmetries are hidden symmetries which become manifest by performing
a reduction of a given discrete system into an effective lower-dimensional one. 
We show how latent symmetries can be leveraged for continuous wave setups in the form of acoustic networks.
These are systematically designed to possess latent-symmetry induced point-wise amplitude parity between selected waveguide junctions
for all low frequency eigenmodes. We develop a modular principle to interconnect
latently symmetric networks to feature multiple latently symmetric junction pairs.
By connecting such networks to a mirror symmetric subsystem, we design asymmetric setups 
featuring eigenmodes with domain-wise parity. Bridging the gap between discrete and continuous models,
our work takes a pivotal step towards exploiting hidden geometrical symmetries in realistic wave setups.
	\end{abstract}
	
	\maketitle
	
	\PRLSec{Introduction}
Symmetries dictate the appearance of fundamental physical laws and allow to make detailed predictions
without solving the underlying equations of motion 
\cite{Schwichtenberg2018PhysicsSymmetry,Peskin2018IntroductionQuantumFieldTheory}.
Selection rules for atoms and molecules 
\cite{Wilson1980MolecularVibrationsTheoryInfrared,Dresselhaus2008GroupTheoryApplicationPhysics},
the emergence of Bloch states and band structures in crystals \cite{NeilW.Ashcroft1976SolidStatePhysics},
and the explanation of spectral degeneracies \cite{Dresselhaus2008GroupTheoryApplicationPhysics}
are all examples for the importance of symmetries.

Recently the concept of latent symmetry has been introduced, that is, a symmetry not of an original Hamiltonian,
but of an equivalent dimensionally reduced effective Hamiltonian \cite{Smith2019PA514855HiddenSymmetriesRealTheoretical}.
Importantly, the presence of a latent symmetry leaves its fingerprints in the original eigenvectors, thereby showing e.g. a parity symmetry of certain eigenvector components only.
This concept has proven fruitful in many different areas such as
the analysis of complex networks 
\cite{Smith2019PA514855HiddenSymmetriesRealTheoretical,Bunimovich2019AMNS4231FindingHiddenStructuresHierarchies,Bunimovich2014IsospectralTransformationsNewApproach}, the explanation of a class of accidental degeneracies 
\cite{Rontgen2021PRL126180601LatentSymmetryInducedDegeneracies}, and can be used to design lattices
with flat bands \cite{Morfonios2021PRB104035105FlatBandsLatentSymmetry}, a topic of major 
current interest \cite{Leykam2018AP31473052ArtificialFlatBandSystems,Leykam2018AP3070901PerspectivePhotonicFlatbands}.
Besides this, a certain subclass of latent symmetries can be closely linked to the graph-theoretical concept of
cospectrality \cite{Kempton2020LAIA594226CharacterizingCospectralVerticesIsospectral,Schwenk1973PTAAC257AlmostAllTreesAre,Godsil2017A1StronglyCospectralVertices}, which is of importance in the context of (almost) perfect state transfer \cite{Bose2003PRL91207901QuantumCommunicationUnmodulatedSpin,Christandl2004PRL92187902PerfectStateTransferQuantum,Plenio2004NJP636DynamicsManipulationEntanglementCoupled,Vinet2012PRA86052319AlmostPerfectStateTransfer,Godsil2012DM312129StateTransferGraphs}.
	
The theory of latent symmetries has so far been developed and applied only to discrete systems.
In this letter, we take the conceptual step of extending and applying the concept of latent symmetry
to a continuous system. We systematically design networks with point-wise amplitude parity between
selected waveguide junctions for all low frequency eigenmodes. Our construction principle yields
asymmetric setups which possess eigenmodes with domain-wise parity.

	\PRLSec{Setup}
	We investigate the eigenmodes of acoustic networks described by the 
	$3D$-Helmholtz equation
	\begin{equation} \label{eq:HelmholtzEq}
	\Delta p + k^2 p = 0
	\end{equation}
	with Neumann hard boundary (wall) conditions on the rigid surfaces of waveguides or cavities, and with $p(\mathbf{r})$ denoting the acoustic pressure field.
	For simplicity, we consider the case where all structures possess the same thickness $d$, so that \cref{eq:HelmholtzEq} can be separated into a $2D$ (x-y plane) and a $1D$ (z-axis) problem.
	
	We begin with networks formed by interconnecting waveguides of equal length $L$.
	If such a network is spatially mirror-symmetric, its eigenmodes have odd or even parity under the reflection, that is, 
	$p(\mathcal{R}(\mathbf{r})) = \pm p(\mathbf{r})$ for all points $\mathbf{r}$,
	with $\mathcal{R}(\mathbf{r})$ denoting the reflection operation.
	In contrast to that, we design asymmetric networks that feature \emph{point-wise parity} in their low-frequency eigenmodes, that is, $p(\mathbf{r}_{n}) = \pm p(\mathbf{r}_{m})$ only for specific locations $\mathbf{r}_{n}, \mathbf{r}_{m}$.
	To reach this goal, we design the network to feature a latent symmetry by tuning the waveguide widths.
	In a next step, we show that those networks can be easily augmented by mirror-symmetric cavity subsystems such that (i) the coupled system features no geometrical symmetry while (ii) the low-frequency eigenmodes have \emph{domain-wise parity}, that is, definite parity everywhere in the cavities.
	
	\PRLSec{Latent symmetries in eigenvalue problems}
	Let us start by sketching the theory of latent symmetries \cite{Smith2019PA514855HiddenSymmetriesRealTheoretical}.
	It is based on the ordinary eigenvalue problem 
	\begin{equation} \label{eq:normalEVP}
	\ham\hspace{0.1em} \evecC = \lambda \evecC
	\end{equation}
	with $\hC$ denoting the Hamiltonian represented by a hermitian matrix.
	To define latent symmetries, we first partition the underlying setup into two subsystems, $S$ and its complement $\sbar$, and write \cref{eq:normalEVP} in block-form as
	\begin{equation}
		\begin{pmatrix}
		\ham_{SS} & \ham_{\ssb} \\
		\ham_{\sbs} & \ham_{\sbb}
		\end{pmatrix} \begin{pmatrix}
		\evecCComponent{S} \\
		\evecCComponent{\sbar}
		\end{pmatrix} = \lambda \begin{pmatrix}
		\evecCComponent{S} \\
		\evecCComponent{\sbar}
		\end{pmatrix} \, .
	\end{equation}
	Assuming for simplicity that $\lambda \, \mathbb{1} - \ham_{\sbb}$ is invertible for any eigenvalue of $\ham$, we can formally solve the second equation for $\evecCComponent{\sbar}$ and insert it into the first.
	This gives us the non-linear eigenvalue problem
	\begin{equation} \label{eq:reducedEVP}
		\ISR{S}{\ham}\hspace{0.1em} \evecCComponent{S} = \lambda \hspace{0.1em} \evecCComponent{S}
	\end{equation}
	with the effective Hamiltonian
	$\ISR{S}{\ham} = \ham_{SS} + \ham_{\ssb} \left(\lambda\, \mathbb{1} - \ham_{\sbb} \right)^{-1} \ham_{\sbs}$  \cite{Smith2019PA514855HiddenSymmetriesRealTheoretical,Kempton2020LAIA594226CharacterizingCospectralVerticesIsospectral,Grosso2013SolidStatePhysics,Rontgen2021PRL126180601LatentSymmetryInducedDegeneracies}. $\ISR{S}{\ham}$ is known as the ``isospectral reduction'' of $\ham$, since in general its (non-linear) eigenvalues are equal to that of the original Hamiltonian $\ham$.
	
	The effective Hamiltonian may or may not have symmetries. If it does, that is, if $\left[\ISR{S}{\ham},M \right] = 0$ for all $\lambda$, where the normal matrix $M$ describes the symmetry operation, then the original Hamiltonian $\hC$ is said to be \emph{latently symmetric} in $S$.
	A latent symmetry has a profound impact on the eigenvectors $\evecC$ of $\ham$: Firstly, it follows from \cref{eq:reducedEVP} that the restriction $\evecCComponent{S}$ of $\evecC$ on $S$ must be an eigenvector of $\ISR{S}{\ham}$.
	Now, since $\ISR{S}{\ham}$ is $M$-symmetric, it follows that (assuming no degeneracies) $\evecCComponent{S}$ must follow this symmetry as well.
	In other words, $\evecC$ must be locally $M$-symmetric on $S$.
	Thus, by tuning the system to feature a latent symmetry with a specific matrix $M$, it is possible to tailor \emph{local properties} of the eigenvectors.	
	To demonstrate the principle of latent symmetry induced design of local properties, we will here focus on the special case of a latent mirror symmetry (LMS) described by $M \equiv \Sigma = \begin{psmallmatrix}
	0 & 1 \\
	1 & 0
	\end{psmallmatrix}$.
	Since the eigenvalues of $\Sigma$ are $\sigma = \pm 1$, this means that one can choose the eigenvectors of $\hC$ to have definite parity on $S := \{u,v\}$.
	That is, they feature \emph{point-wise parity} \footnote{We note that LMS and point-wise parity are also possible in non-hermitian systems.}.
	
	\begin{figure}[htb] 
		\centering
		\includegraphics[max width=\linewidth]{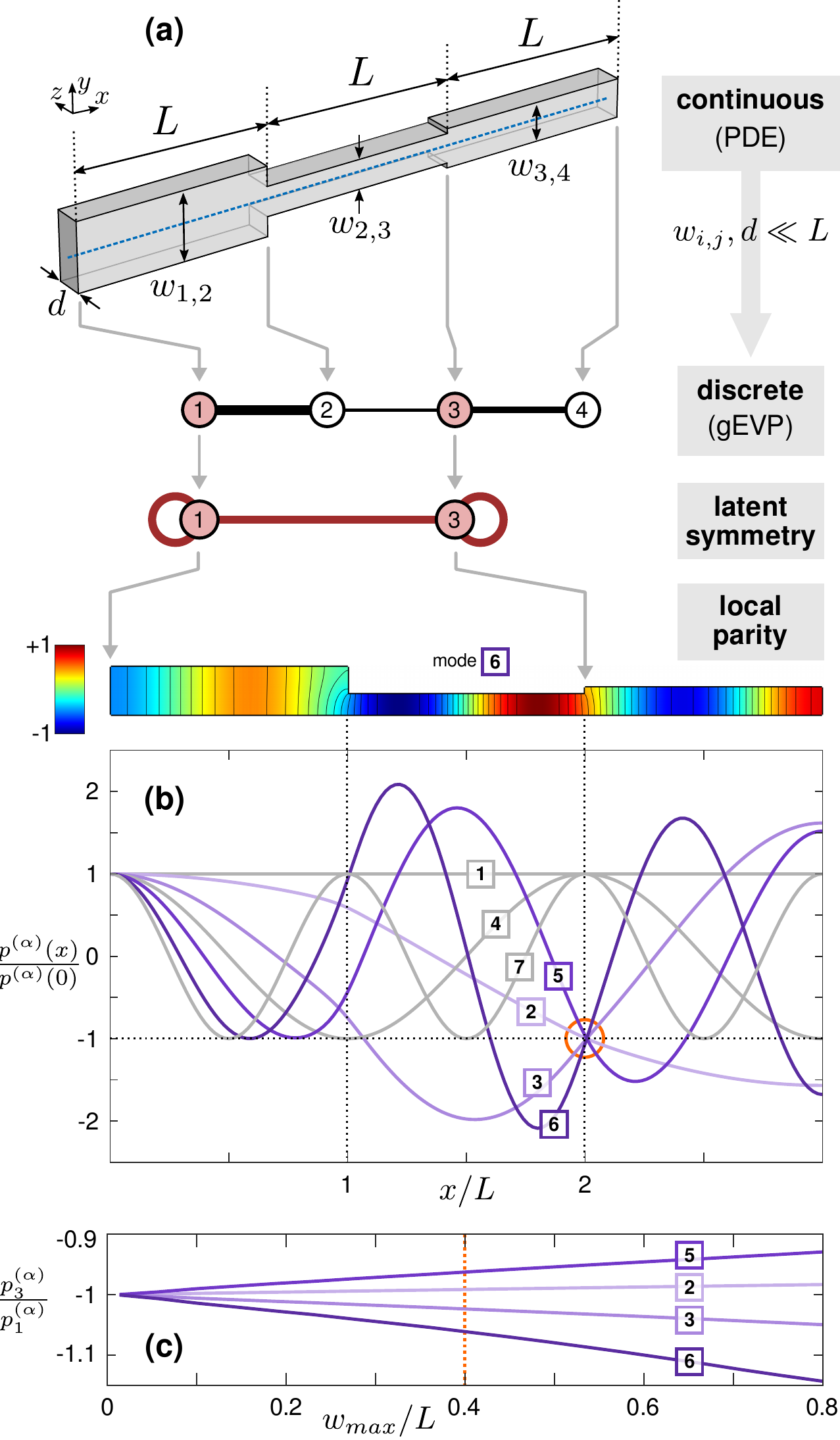}
		\caption{\textbf{(a)} A setup consisting of three waveguides (top) and its mapping to a discrete model (bottom; see text for details). For $w_{1,2} = w_{3,4} + w_{2,3}$, the setup features a latent mirror symmetry with $S = \{1,3\}$, resulting in point-wise parity of the eigenfunctions.
			\textbf{(b)} The amplitude ratio $p^{(\alpha)}(x)/p^{(\alpha)}(0)$ [evaluated along the blue dashed line depicted in (a)] for the first $\alpha=1,\ldots{},7$ eigenmodes of the continuous \cref{eq:HelmholtzEq}.
			\textbf{(c)} The behavior of $p_{3}^{(\alpha)}/p_{1}^{(\alpha)}$ for the non-plane wave modes $\alpha=\{2,3,5,6\}$ for varying $w_{max}/L$ with $w_{max} = max(w_{1,2},w_{2,3},w_{3,4})$.
			For the modes depicted in (b), $w_{max}/L = 0.4$.
		}
		\label{fig:geometryOfSetup}
	\end{figure}
	
	\PRLSec{Latent symmetry in acoustic waveguides}
	In order to construct latently symmetric networks of acoustic waveguides, we focus on narrow waveguides, $w_{n} \ll L$, $d \ll L$, such that the propagation of low-frequency waves in the individual waveguides with width $w_{n}$ and identical thickness $d$ effectively becomes one-dimensional (see Sec. I. of the Supplemental Material for more details).
	In this regime, which we will consider throughout this work, the eigenmode amplitudes at the endpoints of waveguides can be described by the generalized eigenvalue problem (gEVP) \cite{Zheng2019PRA12034014ObservationEdgeWavesTwoDimensional,Coutant2020PRB102214204RobustnessTopologicalCornerModes,Coutant2021JoAP129125108TopologicalTwodimensionalSuSchriefferHeegerAnalog,Coutant2021PRB103224309AcousticSuSchriefferHeegerLatticeDirect}
	\begin{equation} \label{eq:EVP}
	A \evecA = \lambda B \evecA
	\end{equation}
	with $\lambda = \cos(k L)$, $A_{n,m} = w_{n,m}$, and $B$ diagonal with $B_{n,n} = \sum_{m} w_{n,m}$.
	Here, $w_{n,m}$ denotes the width of the waveguide between the endpoints $n$ and $m$, with $w_{n,m} = 0$ if there is no waveguide.
	The eigenvector $\evecA = (p_{1},\ldots{},p_{N})^{T}$ corresponds to the acoustic pressure on the endpoints of the waveguides.
	For our first example, the three-waveguide setup of \cref{fig:geometryOfSetup} (a), the discrete problem is four-dimensional, and we have
	\begin{align}
	A& = \begin{pmatrix}
	0 & w_{1,2} & 0 & 0\\
	w_{1,2}& 0 & w_{2,3}& 0 \\
	0 & w_{2,3} & 0 & w_{3,4} \\
	0 & 0 & w_{3,4} & 0
	\end{pmatrix},\\
	B &= \begin{pmatrix}
	w_{1,2} & 0 & 0 & 0\\
	0& w_{1,2}+w_{2,3} & 0 & 0 \\
	0 & 0 & w_{2,3}+w_{3,4} & 0 \\
	0 & 0 & 0 & w_{3,4}
	\end{pmatrix} \, .
	\end{align}

	Before we continue, we note that the gEVP in the form of \cref{eq:EVP} with $A,B$ real-symmetric and $B$ positive definite is widespread; it occurs in electronic structure models in a non-orthogonal basis in quantum-chemistry \cite{Priyadarshy1996JCP1049473BridgemediatedElectronicInteractionsDifferences}, spring-mass systems, molecular or mechanical vibrations \cite{Wilson1980MolecularVibrationsTheoryInfrared,Shabana2019TheoryVibrationIntroduction}, and it also appears naturally in numerical finite element treatments of wave equations \cite{Bathe1976NumericalMethodsFiniteElement}.
	Now, since the matrix $B$ is positive definite in all these case, we can convert \cref{eq:EVP} to the ordinary symmetric eigenvalue problem \cref{eq:normalEVP} with $\evecC = \sqrt{B}\, \evecA$ and the real-symmetric `Hamiltonian' $\hC = \sqrt{B}^{-1} \, A \, \sqrt{B}^{-1}$.
	This convenient transformation allows us to extend the concept of latent symmetries from ordinary [\cref{eq:normalEVP}] to generalized eigenvalue problems [\cref{eq:EVP}].
		
	Let us now analyze the case where the Hamiltonian $\ham$ corresponding to a gEVP features a LMS for $S := \{u,v\}$.
	Assuming for simplicity that $\hC$ features no degeneracies \footnote{Statements for the case of degeneracies can be found in Sec. II. of the Supplemental Material. The Supplemental Material contains Refs. \cite{Berkolaiko2013186IntroductionQuantumGraphs,Kuchment2002WRM12R1GraphModelsWavesThin,Dalmont1994AA2421LatticesSoundTubesHarmonically,Chung199692SpectralGraphTheory,Eisenberg2019DM3422821PrettyGoodQuantumState,Parlett1998SymmetricEigenvalueProblem,Kempton2020LAIA594226CharacterizingCospectralVerticesIsospectral}}, this latent symmetry induces point-wise parity on $u,v$ onto the eigenvectors $\evecC$ of $\ham$.
	Depending on the structure of $B$, the eigenvectors $\evecA = \sqrt{B}^{-1} \evecC$ of \cref{eq:EVP} then may or may not feature point-wise parity.
	In the special case of acoustic waveguides, however, a LMS of $\hC$ automatically induces point-wise parity, that is, $\evecAComponent{u} = \pm \evecAComponent{v}$ for any eigenvector $\evecA$ (see sections I C and II in the Supplemental Material).
	
	We now apply the concept of latent symmetries to a concrete setup.
	In \cref{fig:geometryOfSetup}, we show a particularly simple waveguide network which, for $w_{1,2} = w_{3,4} + w_{2,3}$, features a LMS for the two junctions $S=\{1,3\}$.
	Thus, the eigenmodes $\evecA$ of the gEVP \cref{eq:EVP} have point-wise parity on $1,3$; the corresponding Hamiltonian has no degeneracy.
	As a consequence, the acoustic pressure at the endpoints $1$ and $3$ [see \cref{fig:geometryOfSetup} (a)] has point-wise parity for low-frequency eigenmodes and narrow waveguides.
	
	Two aspects are noteworthy.
	Firstly, while the Hamiltonian $\ham$ describing \cref{fig:geometryOfSetup} (a) is only four-dimensional, the point-wise parity induced by its LMS has an impact on more than four eigenmodes of the underlying continuous \cref{eq:HelmholtzEq}.
	Indeed,	and as we show in Sec. I. of the Supplemental Material, as long as the low-frequency limit (monomode approximation) is valid, every eigenmode $p$ of \cref{eq:HelmholtzEq} features point-wise parity; this is demonstrated in \cref{fig:geometryOfSetup} (b).
	Secondly, we stress that our theoretical considerations of a latently mirror symmetric waveguide network (LMSWN) are based on approximating \cref{eq:HelmholtzEq} by a \emph{discrete} gEVP.
	Thus, one would expect that our results are valid only in the limiting case of very narrow waveguides.
	The point-wise parity of eigenmodes, however, is robust and it approximately persists even when departing from the limiting case $w_{max}/L \rightarrow 0$, up to roughly $w_{max}/L \simeq 0.2$.
	This is shown in \cref{fig:geometryOfSetup} \textbf{(c)}.
	There, we scale all widths by an identical factor and analyze the deviation from $-1$ of the ratio $p_{1}^{(\alpha)}/p_{3}^{(\alpha)}$ for the eigenmodes $\alpha=\{2,3,5,6\}$ in dependence of this scaling factor.
	For modes $1,4,7$, we note that the point-wise parity is perfect, $p_{1}^{(\alpha)}/p_{3}^{(\alpha)} = 1$, because these modes---for which $k L$ is a multiple of $\pi$---are exact solutions to the PDE of \cref{eq:HelmholtzEq}.
	Irrespective of the individual waveguide widths, these modes do always exist, and as can easily be shown, they exactly fulfill $p_{1}^{(\alpha)}/p_{3}^{(\alpha)} = 1$ even far away from the limit $w_{n,m} \ll L$.
	
	\begin{figure*}[htb] 
		\centering
		\includegraphics[max width=2\columnwidth]{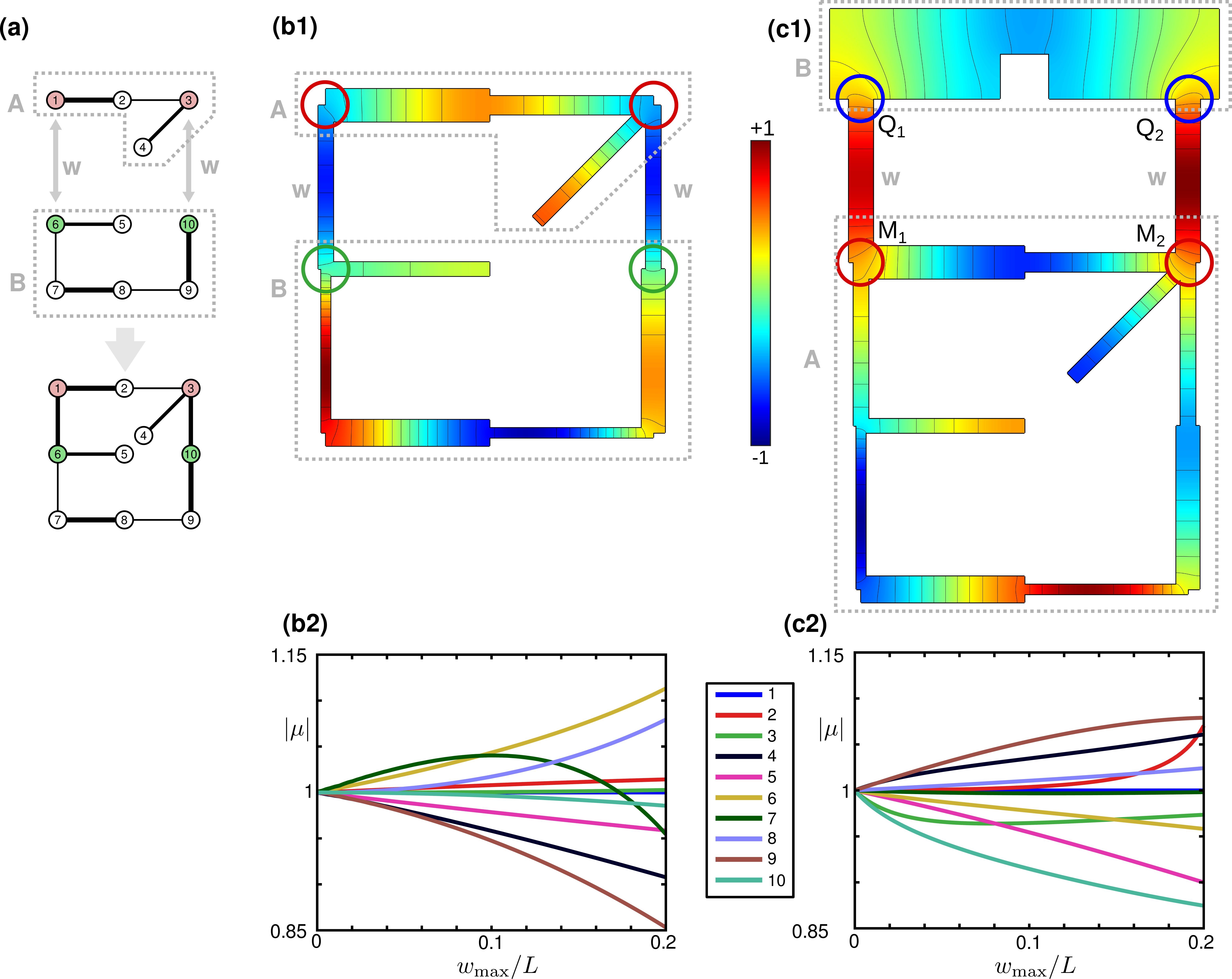}
		\caption{
			\textbf{(a)} Connecting two networks $A$ and $B$ with latent mirror symmetries (in $A$ for $\{1,3\}$, in $B$ for $\{6,10\}$) through the corresponding latent symmetry points (see text for details). For the latent symmetry in $B$, the widths need to be chosen as $w_{5,6} = w_{9,10} - w_{6,7}$ and $w_{8,9} = \frac{w_{6,7} w_{7,8}w_{9,10}}{w_{5,6} w_{7,8}+w_{6,7} w_{9,10}}$.
			\textbf{(b1)} shows the ninth eigenmode of the network constructed by combining $A$ and $B$ in the above manner. The depicted eigenmode features point-wise parity on the junction points $\{1,3\}$ and $\{6,10\}$ (center points of the upper/lower circles, respectively).
			\textbf{(c1)} shows the tenth eigenmode of the system that is obtained by augmenting (b) with a cavity on top.
			This mode features domain-wise parity in the coupled-cavity subsystem.
			\textbf{(b2)} and \textbf{(c2)} show the absolute values of the amplitude ratio $\mu$ (see text) on the center of the red circles in the corresponding setup. For both (b1) and (c1), we have $w_{max}/L = 0.2$.
		}
		\label{fig:niceMode}
	\end{figure*}

	\PRLSec{Network design}
	Having demonstrated a first instance of a LMSWN, let us now address the general construction of such networks.
	This task is equivalent to finding a network geometry with suitable widths $w_{i,j}$ and two sites $S=\{u,v\}$ for which $\ISR{S}{\ham}$ commutes, for all $\lambda$, with $\Sigma= \begin{psmallmatrix}
		0 & 1 \\
		1 & 0
	\end{psmallmatrix}$.
	Expanding $\ISR{S}{\ham}$ into a power series in $\lambda$ shows that this commutation is equivalent to \cite{Kempton2020LAIA594226CharacterizingCospectralVerticesIsospectral,Rontgen2021PRL126180601LatentSymmetryInducedDegeneracies}
	\begin{equation} \label{eq:latSym}
		\left( H^k \right)_{u,u} = \left( H^k \right)_{v,v} \;\forall\; 1 \le k \le N-1 \, ,
	\end{equation}
	with $N$ denoting the dimension of $\ham$.
	Interestingly, an analysis of these $N-1$ conditions on the diagonal elements of the matrix powers of $\ham$ allows to derive generic rules that a LMSWN has to fulfill (see Supplemental Material section IC).
	For example, the sum of widths of waveguides adjacent to $u$ and $v$ must be identical; that is, $B_{u,u} = B_{v,v}$.
	
	The difficulty of finding a latently mirror symmetric configuration clearly depends on the network size and topology.
	In general, for a given $\ham$ of size $N$, the choice of $S$ would be done in principle by trying all $N(N-1)/2$ possible pairs $\{u,v\}$ and testing whether they obey \cref{eq:latSym}.
	For small acoustic networks, as the one in \cref{fig:geometryOfSetup}, a suitable combination of widths and $S$ 
	can be even found analytically yielding the simple condition $w_{1,2} = w_{3,4} + w_{2,3}$; 
	but for larger networks the complexity may become too high.
	Fortunately, there is an alternative means for this problem: As we now demonstrate, smaller LMSWNs can be combined to form \emph{arbitrarily large} networks:
	Given two waveguide networks with latent mirror symmetries for $S_{n} = \{u_{n},v_{n}\}$, $n=1,2$, we connect $u_{1}$ to $u_{2}$ by a narrow waveguide of arbitrary (though small) width $w$, and $v_{1}$ to $v_{2}$ by another waveguide with identical width $w$.
	As shown in Sec. III of the Supplemental Material, the resulting larger network is then guaranteed to feature latent mirror symmetries for both $S_{1}$ and $S_{2}$.
	\Cref{fig:niceMode} (a) demonstrates this principle.
	Here, $A$ denotes the first network, with $S_{1} = \{1,3\}$, which is in fact the three-waveguide network we already encountered in \cref{fig:geometryOfSetup}.
	$B$ denotes a second network of five waveguides which---by finding suitable widths and $S=\{u,v\}$ fulfilling \cref{eq:latSym}---has been designed to feature a LMS for $S_{2} = \{6,10\}$.
	\Cref{fig:niceMode} (b) shows an eigenmode of the resulting setup featuring point-wise parity for all low-frequency eigenmodes \emph{both} on $S_{1}$ and $S_{2}$, as predicted.
	The above principle can be repeated by analogously connecting a third network with $S_{3} = \{u_{3},v_{3}\}$ to either $u_{1},v_{1}$ or $u_{2},v_{2}$.
	A fourth network can then be connected to either of the three $S_{i}$, and so on, ultimately arriving at a modular construction principle.
	
	\PRLSec{Domain-wise parity}
	Instead of coupling two latently symmetric networks, as done in \cref{fig:niceMode} (a) and (b), we could just as well couple a subsystem $B$ with a \emph{conventional} global geometrical symmetry to a latently symmetric network $A$.
	Interestingly, as we now demonstrate in \cref{fig:niceMode} (c), this can even be done when $B$ is no longer a network of thin waveguides but a \emph{spatially extended} setup.
	In that figure, $B$ is an extended, mirror-symmetric cavity, while $A$ corresponds to the setup from \cref{fig:niceMode} (b).
	
	To understand the outcome of this procedure, let us investigate the composite system of the waveguide network (ending at points $M_{1,2}$) and the two waveguides $w$ which end at the two points $Q_{1,2}$.
	Due to latent symmetry, all eigenmodes of this setup have parity, both on $M_{1,2}$ and on $Q_{1,2}$.
	When connecting this composite setup to the extended cavity, the latter ``sees'' only a two-port setup with an impedance relation $\mathbf{p} = Z\, \mathbf{p}'$, where the two-dimensional vectors $\mathbf{p},\mathbf{p}'$ denote the pressure and normal derivative, respectively, at the two points $Q_{1},Q_{2}$.
	Now, as shown in the Supplemental Material, in the low-frequency approximation we have $Z_{11}(k) = Z_{22}(k)$ \emph{for all $k$}.
	As a result, the eigenmodes of the entire interconnected geometry have definite parity in the complete subsystem $B$.
	What is unexpected about this example is that the eigenmodes display this parity even though the geometry of the overall network is not symmetric.
	The \emph{domain-wise} parity observed in \cref{fig:niceMode} (c) is an interesting extension of the other case examples shown in this work, whose eigenmodes featured only point-wise parity.
	
	Similarly to our first setup of \cref{fig:geometryOfSetup}, the observed parity is robust and it remains approximately valid even for the case of waveguides that are not so thin.
	This is demonstrated in \cref{fig:niceMode} (b2) and (c2), where we show the absolute value $|\mu|$ of the pressure ratio $\mu = p_{u}^{(\alpha)}/p_{v}^{(\alpha)}$ for the first $10$ eigenmodes, with $u,v$ denoting the center points of the two red circles in \cref{fig:niceMode} (b1) and (c1).
	
	\PRLSec{Concluding remarks}%
Geometrical symmetries form the basis of regularities and order in wave patterns.
We have demonstrated that point-wise or even domain-wise parities can be systematically introduced in correspondingly asymmetric acoustic networks in their low-frequency eigenmodes.
The origin of this behaviour are hidden or latent symmetries
which can be revealed by an effective Hamiltonian approach.
This constitutes the basis for the design of networks with multiple latent symmetries.
By putting symmetric or anti-symmetric point-sources
where the eigenmodes feature latent-symmetry induced point-wise parity, one may
control the symmetry properties of the wave field.
Within this perspective, when opening up the waveguide network one can imagine to control, e.g., the reflection coefficients of the corresponding multi-port scattering setup in the low-frequency regime \cite{AchilleosIpLatentSymmetryGeneralizedCospectrality}. In a more far reaching perspective latent symmetries might be generalized to PT symmetric or topological waves.

	\begin{acknowledgments}
		The authors are thankful to Vassos Achilleos, Fotios Diakonos, Maxim Pyzh, Olivier Richoux, Georgios Theocharis, and Lasse Wendland for valuable discussions.
	\end{acknowledgments}

\end{document}


\title{Supplemental Material:\\ Hidden symmetries in acoustic wave systems\\\vspace{0.3cm} \small\normalfont by Malte Röntgen, Christian V. Morfonios, Peter Schmelcher, and Vincent Pagneux}
\maketitle

In this supplemental material, mathematical details for the results presented in the main text are included.
\tableofcontents

\section{Latent symmetries in a waveguide network} \label{sec:DiscreteApproximation}
In this section, we present details on the point-wise parity of low-frequency eigenmodes in a waveguide network.
We start by discussing the mapping between the continuous Helmholtz problem and the generalized matrix eigenvalue problem [see also \cite{Berkolaiko2013186IntroductionQuantumGraphs,Coutant2021PRB103224309AcousticSuSchriefferHeegerLatticeDirect}] in \cref{subsec:derivingThegEVP,subsec:ghostModes}. In \cref{sec:constraints}, we investigate the impact that a latent mirror symmetry has on the matrix powers of the Hamiltonian $\ham$ describing the system, and which constraints this puts onto the system itself.
Afterwards, we will discuss latent symmetry in \cref{subsec:latSymInDiscreteWaveguide}.
\Cref{subsec:auxLemma} contains an auxiliary lemma.

\subsection{Deriving the generalized eigenvalue problem in the limit of narrow waveguides} \label{subsec:derivingThegEVP}
In the following, we will show how a network of narrow waveguides of uniform length $L$ can be effectively described by a generalized eigenvalue problem.
Our starting point is the three-dimensional Helmholtz equation
\begin{equation} \label{app:HelmholtzEq}
\Delta p + k^2 p = 0
\end{equation}
with Neumann hard boundary (wall) conditions on the rigid surfaces of waveguides---that is, the normal derivative of $p$ at the surfaces vanishes---, and with $p(\mathbf{r})$ denoting the acoustic pressure field.
For low frequencies and narrow waveguides, that is,  $w_{n} \ll L$, $d \ll L$, with $w_{n}$ denoting the width of the $n$-th waveguide and with $d$ denoting the uniform thickness of the waveguides, \cref{app:HelmholtzEq} effectively becomes one-dimensional, so that in each waveguide we have $p'' + k^{2} p = 0$ \cite{Kuchment2002WRM12R1GraphModelsWavesThin,Coutant2021PRB103224309AcousticSuSchriefferHeegerLatticeDirect,Dalmont1994AA2421LatticesSoundTubesHarmonically}.
Throughout the following, we will always work in this approximation.

\begin{figure}[htb] 
	\centering
	\includegraphics[max width=\linewidth]{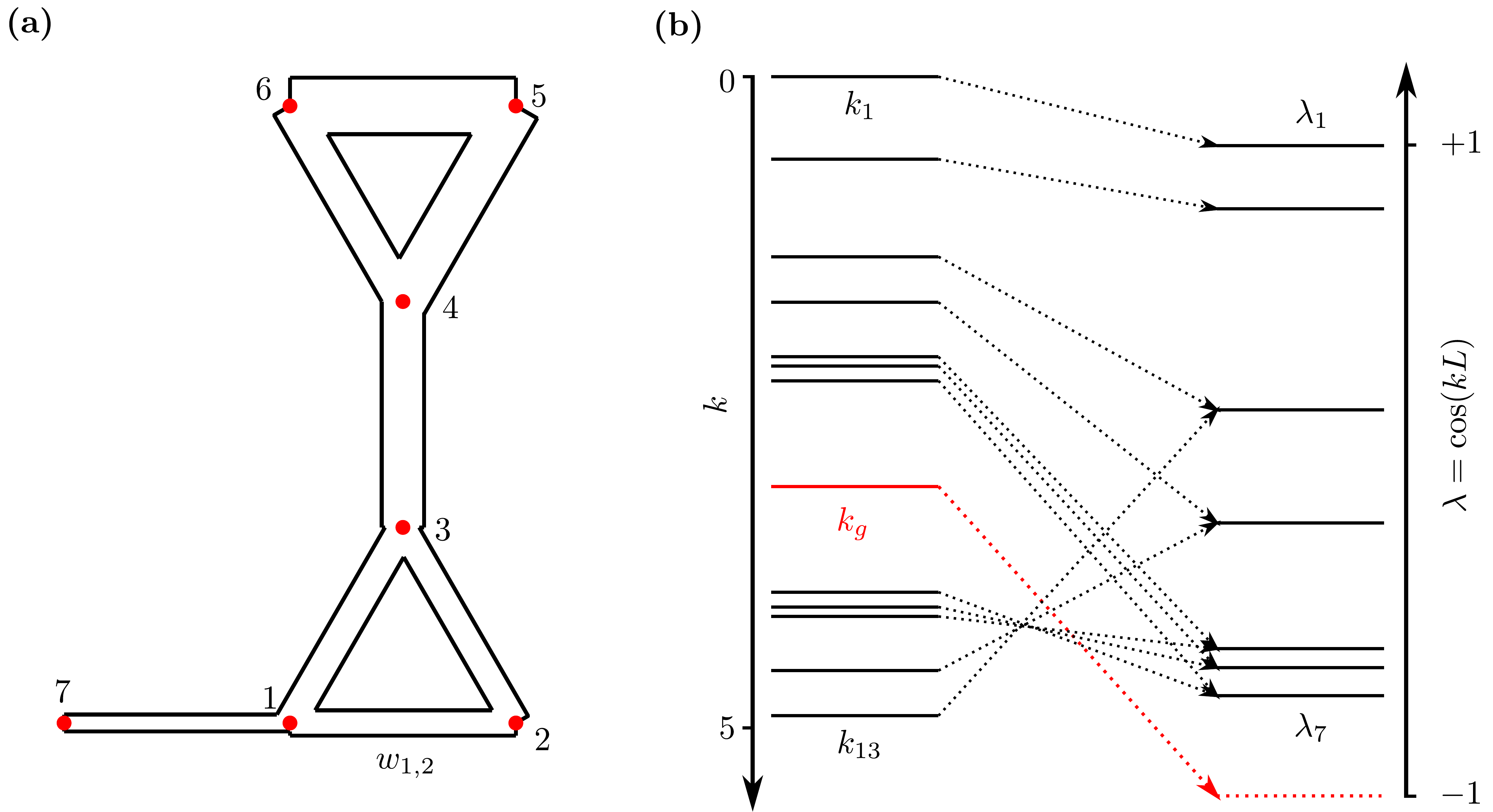}
	\caption{
		\textbf{(a)} An example network of eight waveguides and the labeling of the seven waveguide endpoints in this network (see text for details). \textbf{(b)} Left: Eigenvalue spectrum of the setup shown in (a), computed in the limit of narrow waveguides.
		Only the first $13$ eigenvalues $k_{1},\ldots{},k_{13}$ are shown.
		Right: the eigenvalues $\lambda = \cos(k L)$ of the discrete eigenvalue problem corresponding to the setup in (a).
		Dotted arrows denote the mapping from eigenvalues of the continuous (left) to the discrete (right) model.
		In general, the continuous model might have one or more eigenmodes which vanish on all waveguide endpoints.
		The expression $\cos(k_{g} L)$ of such a ``ghost mode'' not necessarily corresponds to an eigenvalue of the gEVP \cref{app:EVP}, as is the case here (see text for details).
	}
	\label{fig:junctionSketch}
\end{figure}

In order to derive the corresponding generalized eigenvalue problem, we first introduce a set of special points of the waveguide network, and further enumerate them.
These points are (i) junctions of two or more waveguides, and (ii) the isolated extremities of waveguides.
In the following, we will simply call these points ``waveguide endpoints''.
An example for this set of points is given in \cref{fig:junctionSketch} (a), with six junctions (points $1,\ldots{},6$) and one extremity (point $7$).

We now analyze the situation at a point $n$.
This point has one or more neighbors $\mathcal{N}(n)$---that is, the points that are connected to $n$ by a single waveguide---, and as can easily be shown, at the point $n$ the following two criteria must be fulfilled.
(i) the pressure is continuous, and (ii) the fluxes are continuous, that is, $\sum_{m \in \mathcal{N}(n)} A_{n,m} u_{n,m} = 0$.
Here, $A_{n,m} = d\, w_{n,m}$ is the cross-section of the waveguide connecting points $n$ and $m$, and $u_{n,m}$ is proportional to the velocity in this waveguide, with $u_{n,m}$ being oriented towards $n$.
To continue, we note that the pressure fulfills the one-dimensional wave equation $p'' + k^2 p = 0$ in each waveguide.
One can thus use the solution to this second-order differential equation to relate the pressure $p_{m}$ at one of the neighbors $m \in \mathcal{N}(n)$ of $n$ to the pressure $p_{n}$ at $n$ as follows
\begin{equation} \label{app:pressureRelations}
p_{m} = \cos(k L) p_{n} + \frac{\sin(kL)}{k}u_{n,m} \; \forall \; m \in \mathcal{N}(n) \, .
\end{equation}
Multiplying \cref{app:pressureRelations} by $A_{n,m}$ and summing over $m$, we obtain (using the continuity of fluxes)
\begin{equation} \label{eq:protoMatrixEquation}
\sum_{m \in \mathcal{N}(n)} A_{n,m} \, p_{m} = \cos(k L) \sum_{m \in \mathcal{N}(n)} A_{n,m} p_{n} \, .
\end{equation}
Now, by using the convention that $A_{l,l'} = 0$ when there is no waveguide connecting the two points $l,l'$, we can write \cref{eq:protoMatrixEquation} as
\begin{equation}
\sum_{m=1}^{N} A_{n,m} \, p_{m} = \cos(k L) \sum_{m=1}^{N} A_{n,m} p_{n} \, .
\end{equation}

and we see that 
the network can be described by the generalized eigenvalue problem
\begin{equation} \label{app:EVP}
A \evecA = \lambda B \evecA
\end{equation}
with $\lambda = \cos(k L)$, $A_{l,l'} = A_{l,l'}$, and $B$ diagonal with $B_{l,l} = \sum_{l'} A_{l,l'}$.
The eigenvector $\evecA = (p_{1},\ldots{},p_{N})^{T}$ corresponds to the acoustic pressure on the $N$ waveguide endpoints.
In the example of \cref{fig:junctionSketch} (a), we have $N = 7$, with the two matrices $A,B$ given by
\begin{equation}
A = \left(
\begin{array}{ccccccc}
0 & A_{1,2} & A_{1,3} & 0 & 0 & 0 & A_{1,7} \\
A_{1,2} & 0 & A_{2,3} & 0 & 0 & 0 & 0 \\
A_{1,3} & A_{2,3} & 0 & A_{3,4} & 0 & 0 & 0 \\
0 & 0 & A_{3,4} & 0 & A_{4,5} & A_{4,6} & 0 \\
0 & 0 & 0 & A_{4,5} & 0 & A_{5,6} & 0 \\
0 & 0 & 0 & A_{4,6} & A_{5,6} & 0 & 0 \\
A_{1,7} & 0 & 0 & 0 & 0 & 0 & 0 \\
\end{array}
\right)
\end{equation}
and
\begin{equation}
B = \left(
\begin{array}{ccccccc}
A_{1,2}+A_{1,3}+A_{1,7} & 0 & 0 & 0 & 0 & 0 & 0 \\
0 & A_{1,2}+A_{2,3} & 0 & 0 & 0 & 0 & 0 \\
0 & 0 & A_{1,3}+A_{2,3}+A_{3,4} & 0 & 0 & 0 & 0 \\
0 & 0 & 0 & A_{3,4}+A_{4,5}+A_{4,6} & 0 & 0 & 0 \\
0 & 0 & 0 & 0 & A_{4,5}+A_{5,6} & 0 & 0 \\
0 & 0 & 0 & 0 & 0 & A_{4,6}+A_{5,6} & 0 \\
0 & 0 & 0 & 0 & 0 & 0 & A_{1,7} \\
\end{array}
\right) \, ,
\end{equation}
where we remind the reader that $A_{m,n}=d w_{m,n}$ is the cross-sectional area, meaning that matrices $A$ and $B$ are just decided by the geometrical parameters.
For the system in \cref{fig:junctionSketch} (a), the waveguides are of length $L = 1$, with the widths $w_{i,j}$ given by
\begin{equation}
w_{1,2}=\frac{9}{80},\; w_{1,3}=\frac{11}{80},\;
w_{1,7}=\frac{3}{40},\; w_{2,3}=\frac{1}{8},\;
w_{3,4}=\frac{3}{16},\; w_{4,5}=\frac{7}{32},\;
w_{4,6}=\frac{13}{80},\; w_{5,6}=\frac{1}{4}
\end{equation}
To compute the eigenvalues in the limit of small waveguides [as shown in \cref{fig:junctionSketch} (b)], each width was divided by $125$, so that $w_{max}/L = \frac{1}{500}$.

\subsection{The appearance of ghost modes} \label{subsec:ghostModes}
From the above, it follows that---in the limit of narrow waveguides---the vector $\evecA = (p_{1},\ldots{},p_{N})^{T}$ built from an eigenmode $p$ of the continuous problem must fulfill the gEVP \cref{app:EVP}.
Thus, whenever $p$ does not identically vanish on all waveguide endpoints, $\evecA$ is an eigenvector of the gEVP with eigenvalue $\lambda = \cos(k L)$, with $k$ defined by the Helmholtz equation $\Delta p + k^2 p = 0$ which the eigenmode $p$ fulfills.

On the other hand, when $p$ identically vanishes on all waveguide endpoints, we have
\begin{equation} \label{eq:ghostModes}
\evecA = \mathbf{0} = (0,\ldots{},0)^{T} \, .
\end{equation}
Thus, since $\evecA$ is the zero vector, and although it fulfills \cref{app:EVP}, it is by definition \emph{not an eigenvector} of this equation.
We stress that the eigenmode $p_{g}$ does not vanish everywhere in the continuous setup, but since it has nodes on all waveguide endpoints, it is ``invisible'' for the discrete model.
Consequently, we call $p_{g}$ a ``ghost mode''.
We note that ghost modes are rather common.
Indeed, even the simple setup depicted in \cref{fig:junctionSketch} features such a mode.

In this respect, two things are noteworthy.
Firstly, any ghost mode must fulfill $\sin(k_{g} L) = 0$, or, equivalently, $\cos(k_{g} L) = \pm 1$, as can be easily seen from solving the one-dimensional Helmholtz equation $p_{g}'' + k_{g}^2 p_{g} = 0$ in each waveguide and demanding that $p_{g} = 0$ on all waveguide endpoints.
Secondly, since the ghost mode is invisible for the discrete model, the value $\cos(k_{g} L)$ may or may not be an eigenvalue of the gEVP \cref{app:EVP}.
An example for the case where $\cos(k_{g} L) = -1$ is not contained in the eigenvalue spectrum of the gEVP is shown in \cref{fig:junctionSketch} (b).
We stress that, even in the case where $\cos(k_{g} L)$ is an eigenvalue of the gEVP, the ghost mode is still ``invisible'' for the discrete model.
What happens in this case is that there exists another, non-ghost mode $p_{ng}$ fulfilling $\cos(k_{ng} L ) = \cos(k_{g} L)$.

\subsection{Constraints imposed by latent symmetry on a waveguide network} \label{sec:constraints}

Let us now derive some conditions that a waveguide network with a latent reflection symmetry for the junctions $u,v$ has to fulfill.
In order to feature such a symmetry, the Hamiltonian $H$ describing this network must fulfill equation (8) of the manuscript, that is,
\begin{equation} \label{eq:latSym}
\left( H^k \right)_{u,u} = \left( H^k \right)_{v,v} \;\forall\; 1 \le k \le N-1 \, ,
\end{equation}
where $N$ is the dimension of $H$ (that is, the number of junctions). 

If \cref{eq:latSym} holds for a Hamiltonian describing a waveguide network, then the eigenvectors $\evecC$ of $\ham$ have point-wise parity on $u,v$ (if there are degeneracies, the eigenvectors can be chosen to have such parity).
This allows us to derive an essential condition that our waveguide network has to fulfill, as we now show.
To this end, we start by noting that one can easily show that the constant vector $\mathbf{1} := (1,\ldots{},1)^T$ is a \emph{non-degenerate} eigenvector of the gEVP $A \evecA = \lambda B \evecA$ with eigenvalue $\lambda = 1$.
Now, since $\sqrt{B} \, \mathbf{1}$ is an eigenvector of $\ham$ (to the same eigenvalue), it must have point-wise parity on $u,v$. It follows that $B_{u,u} = B_{v,v}$.
Now, by construction, $B$ is a diagonal matrix, with the entry $B_{u,u} = \sum_{i \in \mathcal{N}(u)} w_{i,u}$ where $\mathcal{N}(u)$ are the neighbors of $u$, that is, the junctions that are connected to the junction $u$ through a waveguide (of width $w_{i,u}$).
Thus, we have
\begin{equation}
\sum_{i \in \mathcal{N}(u)} w_{i,u} = \sum_{i \in \mathcal{N}(v)} w_{i,v} \, .
\end{equation}
In words, \emph{the total width of waveguides connected to junction $u$ must be equal to the total width of waveguides connected to junction $v$}.
In the special case of identical waveguides, $w_{i,j} \equiv w$, we see that $u$ and $v$ have to have the same number of neighbors.

A second requirement on the waveguide network can be derived from evaluating \cref{eq:latSym} for $k=2$. Assuming that $B_{u,u} = B_{v,v}$, this equation reads
\begin{equation}
\sum_{i \in \mathcal{N}(u)} \frac{w_{u,i}^2}{B_{i,i}} = \sum_{i \in \mathcal{N}(v)} \frac{w_{v,i}^2}{B_{i,i}} \, .
\end{equation}
In the special case of uniform waveguides, we get
\begin{equation}
\sum_{i \in \mathcal{N}(u)} \frac{1}{B_{i,i}} = \sum_{i \in \mathcal{N}(v)} \frac{1}{B_{i,i}} \; \Rightarrow \; \sum_{i \in \mathcal{N}(u)} \frac{1}{N(i)} = \sum_{i \in \mathcal{N}(v)} \frac{1}{N(i)}
\end{equation}
where $N(i)$ is the number of neighbors of the junction $i$.
In the special case where $N(u) = N(v) = 1$, we see that $u,v$ have to have \emph{the same number of next-neighbors}.

In a similar manner, one can use \cref{eq:latSym} for higher values of $k$ to derive more statements/constraints on the topology and the waveguide widths of the waveguide network.

\subsection{Parity of all eigenmodes for a latently symmetric waveguide network}\label{subsec:latSymInDiscreteWaveguide}	
In the above, we have investigated the mapping between the continuous Helmholtz equation and the gEVP in detail.
This allows us to make the following statement:

\emph{Let $\subsystem{}$ be a waveguide network, with all waveguides having the same length $L$.
	Let $A,B$ be the $N$-dimensional matrices used in the gEVP \cref{app:EVP} corresponding to this network, and let $u,v$ be latently mirror symmetric, that is, $\left( H^k \right)_{u,u} = \left( H^k \right)_{v,v} \;\forall\; 1 \le k \le N-1 \,$.
	Then, in the limit of narrow waveguides, all eigenmodes $\{p\}$ of the underlying continuous problem can be chosen to be (i) pairwise orthogonal, and (ii) to have parity at the points corresponding to $u$ and $v$.}

To see this, we start by noting that any non-degenerate eigenmode $p_{s}$ (where the ``s'' stands for solitary) has to have parity on the waveguide endpoints $u,v$.
That is, $p_{s}(\mathbf{u}) = \pm p_{s}(\mathbf{v})$, where $\mathbf{u},\mathbf{v}$ are the position vectors corresponding to the waveguide endpoints $u,v$.
This statement can be easily shown by considering the following two cases.
If $p_{s}$ is a ghost mode, then it vanishes on all waveguide endpoints; in particular, it has parity on $u,v$.
If, on the other hand, $p_{s}$ is not a ghost mode, then one can easily show that the corresponding eigenvector $\evecA$ of the gEVP is non-degenerate.
Thus, $\evecA$ has to have parity, and so does $p_{s}$.
Thus, it suffices to investigate only degenerate eigenmodes.
To this end, let $\{p^{(1)},\ldots{},p^{(n)}\}$ be a set of $n$ degenerate and pairwise orthogonal eigenmodes of the continuous problem, with common eigenvalue $k$.
From \cref{subsec:derivingThegEVP,subsec:ghostModes}, we see that the set $\{p^{(1)},\ldots{},p^{(n)}\}$ of continuous eigenmodes corresponds to a set of $n$ vectors $\{\evecA^{(1)},\ldots{},\evecA^{(n)}\}$, each of them fulfilling $A \evecA = \lambda B \evecA$ with $\lambda = \cos(k L)$.
We then distinguish two cases:
\begin{enumerate}
	\item $\sin(k L) \ne 0$\\
	As can easily be shown, due to $\sin(k L) \ne 0$ none of the $\{\evecA^{(1)},\ldots{},\evecA^{(n)}\}$ is the zero vector $\mathbf{0} = (0,\ldots{},0)^{T}$.
	Thus, $\{\evecA^{(1)},\ldots{},\evecA^{(n)}\}$ are eigenvectors of the gEVP $A \evecA = \lambda B \evecA$ with common eigenvalue $\lambda = \cos(k L)$.
	Moreover, as we show in the next \cref{subsec:auxLemma}, the $\{\evecA^{(1)},\ldots{},\evecA^{(n)}\}$ are pairwise $B$-orthogonal.
	
	As a consequence of \cref{thm:DefinitePartityConditions} of \cref{sec:DefinitePartityConditions} below, we can superpose the $\{\evecA^{(1)},\ldots{},\evecA^{(n)}\}$ such that the resulting set $\{\widetilde{\evecA}^{(1)},\ldots{},\widetilde{\evecA}^{(n)}\}$ is (i) pairwise $B$-orthogonal and (ii) has parity on sites $u$ and $v$.
	It is straightforward to show that this means that one can superpose the continuous eigenmodes such that the resulting superpositions $\{\tilde{p}^{(1)},\ldots{},\tilde{p}^{(n)}\}$ are (i) pairwise orthogonal and (ii) have parity at the points corresponding to $u$ and $v$.
	
	\item $\sin(k L) = 0$\\
	There are two possible cases to consider.
	In the first case, all $\evecA^{(i)} = \mathbf{0}$ are zero.
	We thus trivially have $\evecAComponent{u}^{(i)} = \evecAComponent{v}^{(i)}$ for all $i$, so that the eigenmodes $\{p^{(1)},\ldots{},p^{(n)}\}$ have parity on the waveguide endpoints $u$ and $v$.
	
	In the second case, at least one of the $\evecA^{(i)} \ne \mathbf{0}$ (in the following denoted by $\mathbf{x}$), and from the above we see that this vector is an eigenvector of the gEVP to eigenvalue $\lambda = \cos(k L)$.
	Since $\sin(k L) = 0$, we thus have $\lambda = \pm 1$.
	As can easily be shown (along the lines of, e.g., Chapter 1 of \cite{Chung199692SpectralGraphTheory}), $\mathbf{x}$ is a \emph{non-degenerate} eigenvector of the gEVP.
	This non-degeneracy has two consequences.
	Firstly, it follows that $\mathbf{x}$ has parity on $u$ and $v$.
	Secondly, we see that each of the other $n-1$ continuous eigenmodes either (i) identically vanishes on all waveguide endpoints, and thus in particular on $u$ and $v$, or (ii) its projection to the waveguide endpoints is parallel to $\mathbf{x}$.
	In both cases, we see that the $\{p^{(1)},\ldots{},p^{(n)}\}$ have parity on the waveguide endpoints $u$ and $v$.
\end{enumerate}

\subsection{Auxillary Lemma on orthogonality} \label{subsec:auxLemma}
Let $p,q$ denote two degenerate eigenmodes of the $3D$-Helmholtz equation \cref{app:HelmholtzEq}, that is, of $\Delta p + k^2 p = 0$ with eigenvalue $k$ and such that $\sin(k L) \ne 0$.
Moreover, let $\mathbf{P},\mathbf{Q}$ denote the $N$-dimensional vectors obtained from $p,q$ by taking their amplitude at the $N$ waveguide endpoints.
In the following, we will consider the limit of narrow waveguides, and relate the overlap integral $\int p^{*}\, q\, dV$ (star denoting complex conjugate) to an expression involving $\mathbf{P}^{\dagger} B \mathbf{Q}$, with $B$ as above.
In particular, we will show that $p,q$ are orthogonal if and only if $\mathbf{P}^{\dagger} B \mathbf{Q} = 0$, that is, if and only if $\mathbf{P},\mathbf{Q}$ are $B$-orthogonal.

In this limit, in the waveguide $W_{i,j}$ connecting endpoints $i$ and $j$, the overlap integral becomes
\begin{equation} \label{eq:singleOverlapIntegral}
\int_{V=W_{i,j}} p^{*}\, q\, dV = A_{i,j} \int_{0}^{L} p^{*}(x) q(x) dx = A_{i,j} \Big(\alpha  \left(P_i^{*}
Q_i + P_j^{*} Q_j\right) + \beta \left(P_{j}^{*} Q_i + P_{i}^{*} Q_j\right) \Big)
\end{equation}
where $A_{i,j}$ denotes the cross-section of the waveguide $W_{i,j}$, $P_{l},Q_{l}$ denote the function values of $p,q$ at vertex $l$, respectively, and with
\begin{equation}
\alpha = \frac{1}{2} \left(\frac{L}{\sin (k
	L)^2}-\frac{1}{k \tan (k
	L)}\right), \quad \beta = \frac{1-\frac{k L}{\tan (k L)}}{2 k
	\sin (k L)} \, .
\end{equation}

When computing the total integral $\int_{V} p^{*} q dV$ over the whole waveguide setup, we obtain
\begin{equation}
\int_{V} p^{*} q dV =\alpha \sum_{i} \sum_{j=\mathcal{N}(i)}  A_{i,j} P_{i}^{*} Q_{i}  + \beta \sum_{i} \sum_{j=\mathcal{N}(i)} A_{i,j} P_{i}^{*} Q_{j}\, .
\end{equation}
We can then simplify the second summand using \cref{eq:protoMatrixEquation}, which---in the notation of \cref{eq:singleOverlapIntegral}---reads for the mode $q$
\begin{equation}
\sum_{j \in \mathcal{N}(i)} A_{i,j} \, Q_{j} = \cos(k L) \sum_{j \in \mathcal{N}(i)} A_{i,j} Q_{i}\, ,
\end{equation}
yielding
\begin{align}
\int_{V} p^{*} q dV &= \alpha \sum_{i} \sum_{j=\mathcal{N}(i)} A_{i,j} P_{i}^{*} Q_{i}  + \cos(k L) \beta \sum_{i} \sum_{j=\mathcal{N}(i)} A_{i,j}  P_{i}^{*} Q_{i}\\
&= \Big(\alpha + \cos(k L) \beta \Big) \sum_{i} \sum_{j=\mathcal{N}(i)}  A_{i,j} P_{i}^{*} Q_{i} \\
&= \Big(\alpha + \cos(k L) \beta \Big) \mathbf{P}^{\dagger} B \mathbf{Q} \\
&= \frac{L}{2} \mathbf{P}^{\dagger} B \mathbf{Q}\,.
\end{align}
Thus, we see that the two modes $p,q$ are orthogonal if and only if their discrete counterparts $P,Q$ are $B$-orthogonal, as claimed.

\section{Definite parity of eigenvectors for block-diagonal $B$} \label{sec:DefinitePartityConditions}
In the main manuscript, we have discussed latent symmetries in the generalized eigenvalue problem $A \evecA = \lambda B \evecA$ with $A,B$ real symmetric and $B$ positive definite.
In particular, we analyzed the case where two sites $S = \{u,v\}$ are latently mirror symmetric in the real-symmetric `Hamiltonian' $\hC = \sqrt{B}^{-1} \, A \, \sqrt{B}^{-1}$.
That is, for any non-negative integer $k$,
\begin{equation} \label{eq:latSym2}
	\left(\ham^{k} \right)_{u,u} = \left(\ham^{k} \right)_{v,v} \, .
\end{equation}

In the following, we will analyze the impact that such a latent mirror symmetry of $\ham$ has on the eigenvectors $\evecA$ occurring in $A \evecA = \lambda B \evecA$ in more detail.

\begin{Theorem}{}{DefinitePartityConditions} 
	Let $A,B$ real symmetric and $B$ positive definite, and $S = \{u,v\}$ be latently mirror symmetric in $\hC = \sqrt{B}^{-1} A \sqrt{B}^{-1}$, and let us denote two eigenvectors $\evecA,\evecA'$ as $B$-orthogonal if $\evecA^{\dagger}B\evecA' = 0$ when $\evecA \ne \evecA'$.
	Then, for block-diagonal $B = B_{SS} \oplus B_{\sbb}$, the $B$-orthogonal eigenvectors $\evecA$ of the gEVP $A \evecA = \lambda B \evecA$ can be chosen to all have definite parity on $u$ and $v$ if and only if $B_{u,u} = B_{v,v}$.
\end{Theorem}
\begin{proof}
	Latent mirror symmetry of $\{u,v\}$ is equivalent to
	\begin{equation}
	\left(\hC^{k} \right)_{u,u} = \left(\hC^{k} \right)_{v,v}\;\forall\;k \,.
	\end{equation}
	Since $B$ is block-diagonal and positive definite, one can uniquely define its square root as the positive definite matrix 
	\begin{equation}
	B_{SS}^{-1/2} = \left(B_{SS}^{-1} \right)^{1/2} = \left(\frac{1}{|B_{SS}|} \begin{pmatrix}
	B_{v,v} & -B_{u,v} \\
	-B_{u,v} & B_{u,u}
	\end{pmatrix} \right)^{1/2} = \frac{1}{t |B_{SS}|^{1/2}} \begin{pmatrix}
	B_{v,v} + s & -B_{u,v} \\
	-B_{u,v} & B_{u,u} + s
	\end{pmatrix}
	\end{equation}
	where $t = \sqrt{Tr(B_{SS}) + 2s}$ with $s = \sqrt{|B_{SS}|}$.
	
	Since $\hC$ is real and symmetric, we can use its latent symmetry to construct its eigenvectors by means of Lemma 2.5 of Ref. \cite{Eisenberg2019DM3422821PrettyGoodQuantumState}.
	That is, for each $d$-fold degenerate eigenvalue $\lambda$, we can choose the corresponding eigenvectors $\evecC^{(1)},\ldots{},\evecC^{(d)}$ to be pairwise orthonormal and such that
	\begin{itemize}
		\item there is at most one eigenvector $\evecC^{+}$ fulfilling $\evecCComponent{u}^{+} = \evecCComponent{v}^{+} \ne 0$.
		\item there is at most one eigenvector $\evecC^{-}$ fulfilling $\evecCComponent{u}^{-} = -\evecCComponent{v}^{-} \ne 0$.
		\item all remaining $n$ eigenvectors to this eigenvalue vanish on $u$ and $v$, that is, $\evecCComponent{u}^{(0,i)} = \evecCComponent{v}^{(0,i)} = 0$ for $i=1,\ldots{},n$.
	\end{itemize}
	We note that it can be easily shown---see, for instance, Theorem 4 in the Supplemental Material of \cite{Rontgen2021PRL126180601LatentSymmetryInducedDegeneracies}---that $\evecC^{(1)},\ldots{},\evecC^{(d)}$ are real.
	
	We then construct a corresponding basis of eigenvectors $\evecA$ of the gEVP by multiplying each $\evecC$ from the left with $\sqrt{B}^{-1}$.
	As can be directly shown, these eigenvectors $\evecA$ are pairwise $B$-orthogonal, that is, $\evecA^{\dagger}B\evecA' = 0$ when $\evecA \ne \evecA'$.
	We note that this $B$-orthogonality is the correct generalization of ``normal'' orthogonality $\evecC^{\dagger} \evecC' = 0$ for $\evecC \ne \evecC'$ (used for hermitian matrices) to the gEVP (see, e.g., Theorem 15.3.3 in Ref. \cite{Parlett1998SymmetricEigenvalueProblem}).
	
	\begin{itemize}
		\item[``$\Rightarrow$''] Since $B$ is block-diagonal, the $S$-components of $\evecA= \sqrt{B}^{-1} \evecC$ are given by $\evecAComponent{S} = \sqrt{B_{SS}}^{-1} \evecCComponent{S}$.
		We thus obtain 
		\begin{align} \label{eq:transformedEigenvectors1}
		\sqrt{B_{SS}}^{-1} \evecCComponent{S}^{(0,i)} &= 0 \;\forall\; i\\
		\sqrt{B_{SS}}^{-1} \evecCComponent{S}^{\pm} &= \frac{1}{t |B_{SS}|^{1/2}} a \begin{pmatrix}
		s + B_{v,v} \mp B_{u,v} \\
		\pm(s + B_{u,u} \mp B_{u,v})
		\end{pmatrix}\label{eq:transformedEigenvectors2}
		\end{align}
		where $a \ne 0$ denotes the amplitude of $\evecC^{(\pm)}$ on $u$.
		We thus obtain
		$\evecAComponent{u} = \alpha_{\pm} \evecAComponent{v}$,
		with the scaling factors
		\begin{equation} \label{eq:scalingRelations}
		\alpha_{\pm} = \frac{\pm s \pm B_{u,u} - B_{u,v}}{s \mp B_{u,v} + B_{v,v}}
		\end{equation}
		For $B_{u,u} = B_{v,v}$ we obtain $\evecAComponent{u} = \pm \evecAComponent{v}$.
		\item[``$\Leftarrow$'']
		We need to distinguish two cases.
		If the gEVP features no degeneracies, then the eigenvectors \cref{eq:transformedEigenvectors1,eq:transformedEigenvectors2} are, up to global phase, unique, and one sees that they have no parity.
		If, however, the gEVP features degeneracies, then the eigenvectors \cref{eq:transformedEigenvectors1,eq:transformedEigenvectors2} are just a particular choice, and it could in principle be that one can superpose degenerate eigenvectors $\{ \evecA \}$ such that the resulting superpositions $\{\widetilde{\evecA} \}$ are (i) pairwise $B$-orthogonal and (ii) \emph{all of them} have definite parity on $u,v$.
		As we now show, however, simultaneously fulfilling both (i) and (ii) is impossible, and this proves the theorem.
		
		Firstly, let us note that the only possibility to achieve (ii) for a given eigenvalue $\lambda$ is that there exist \emph{both} positive and negative parity eigenvectors $\evecC^{+}$ and $\evecC^{-}$ to this eigenvalue $\lambda$.
		Since $\hC$ and the gEVP share the same eigenvalue spectrum, instead of superposing $\evecA^{+}$ and $\evecA^{-}$, we can equivalently superpose $\evecC^{\pm}$ 
		\begin{align}
		\widetilde{\evecC}_{+} &= c_{+} \evecC^{+} + c_{-} \evecC^{-} \\
		\widetilde{\evecC}_{-} &= d_{+} \evecC^{+} + d_{-} \evecC^{-} 
		\end{align}
		and (i) and (ii) are equivalent to 
		\begin{align}
		\widetilde{\evecC}_{+}^{\dagger} \widetilde{\evecC}_{-} &= 0 \\
		\sqrt{B_{SS}}^{-1} \widetilde{\evecC}_{\pm} &= \beta \begin{pmatrix}
		1 \\
		\pm 1
		\end{pmatrix}
		\end{align}
		with $\beta \ne 0$.
		As can easily be shown, simultaneously fulfilling these equations is impossible.
	\end{itemize}
\end{proof}
We remark that, as shown in \cref{sec:constraints}, for an\emph{ acoustic waveguide network}, a latent mirror symmetry of $\{u,v\}$ automatically implies $B_{u,u} = B_{v,v}$.

\section{Proof for the modular construction principle of latent mirror symmetries in acoustic waveguides} \label{sec:modPrincipleProof}
In the following, we prove the construction principle for acoustic waveguide networks featuring latently symmetric sites.
\begin{figure}[htb] 
	\centering
	\includegraphics[width=0.4\textwidth]{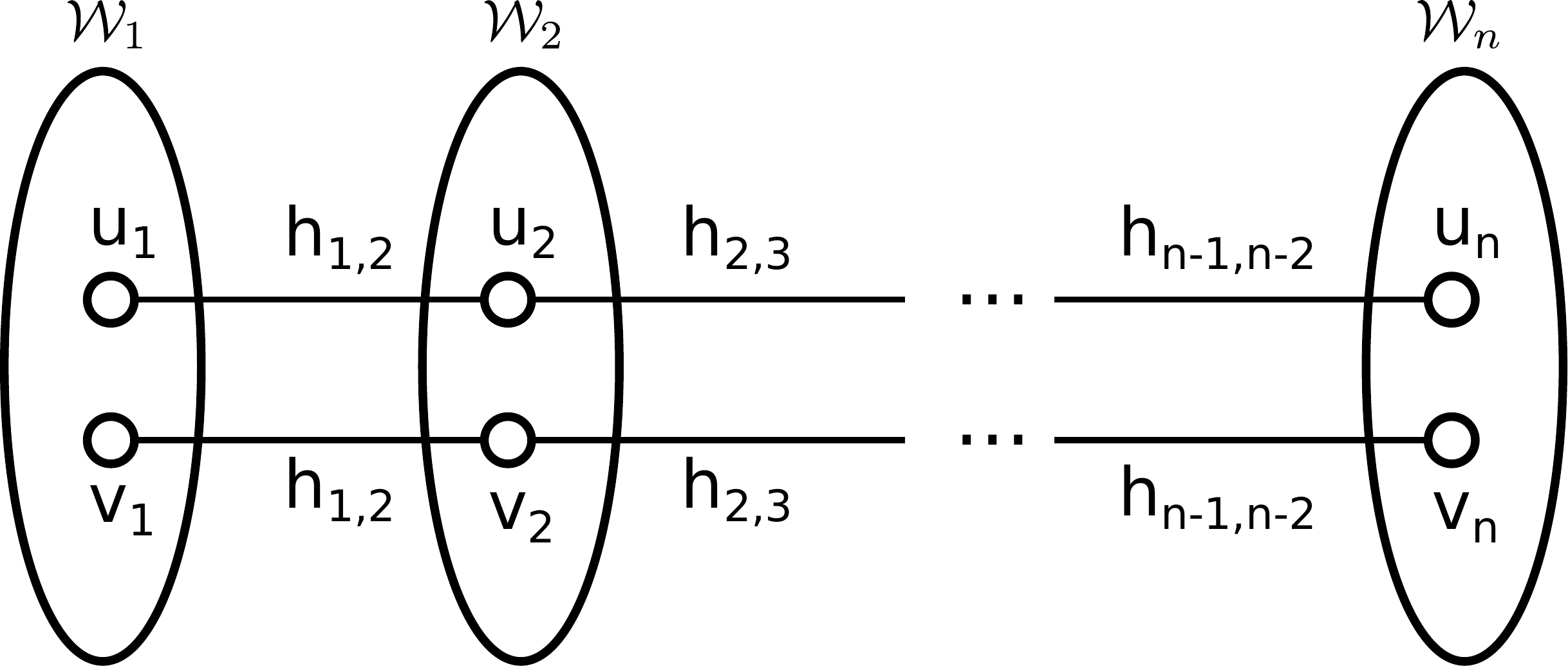}
	\caption{A visualization of the modular construction principle (see text for details).
	}
	\label{fig:interconnectionSketch}
\end{figure}
\begin{Theorem}{}{interconnection}	
	Let $\subsystem{1},\ldots{},\subsystem{n}$ be $n$ disconnected acoustic waveguide networks, with the network $\subsystem{i}$ described by matrices $A_{i}$ and $B_{i}$ as defined in the main text, which in particular implies that $A_{i}$ and $B_{i}$ are real symmetric, with $B_{i}$ additionally being diagonal and positive definite.
	Furthermore, let $S_{i} = \{u_{i},v_{i}\}$ be latently mirror symmetric in $\subsystem{i}$, i.e., $\left(\ham_{i}^{k} \right)_{u,u} = \left(\ham_{i}^{k} \right)_{v,v}$ for all $k$, with $\ham_{i} = \sqrt{B_{i}}^{-1} \, A_{i} \, \sqrt{B_{i}}^{-1}$.
	Then, when connecting the networks as depicted in \cref{fig:interconnectionSketch}, \emph{each} of the $S_{i}$ is latently mirror symmetric in the resulting larger system.
\end{Theorem}
\begin{proof}
	Let us first only look at the network $\subsystem{1} + \subsystem{2}$ obtained by coupling only $\subsystem{1}$ and $\subsystem{2}$ through waveguides of identical cross-section $h_{1,2}$; see \cref{fig:interconnectionSketch} for a sketch of the setup.
	We let $\hC$ denote the matrix derived from the gEVP of $\subsystem{1} + \subsystem{2}$.
	For simplicity, we omit the subscript of $h_{1,2}$.
	
	Denoting the complement of $\sO = \{u_{1},v_{1}\}$ within $\subsystem{1}$ by $\sBO$, and similarly the complement of $\sT= \{u_{2},v_{2}\}$ within $\subsystem{2}$ by $\sBT$, we can write
	\begin{equation}
	A = \begin{pmatrix}
	\comp{A}{\sO}{\sO} & \comp{A}{\sO}{\sBO} & h I & 0 \\
	\comp{A}{\sBO}{\sO} & \comp{A}{\sBO}{\sBO} & 0 & 0\\
	h I & 0 & \comp{A}{\sT}{\sT} & \comp{A}{\sT}{\sBT} \\
	0 & 0 & \comp{A}{\sBT}{\sT} & \comp{A}{\sBT}{\sBT}
	\end{pmatrix}
	\end{equation}
	where $\comp{A}{\sO}{\sO} = \comp{A_{1}}{\sO}{\sO}$, $\comp{A}{\sT}{\sBT} = \comp{A_{2}}{\sT}{\sBT}$, and so on.
	Within this numeration, the square root of the new matrix $B$ for the composite network can be written as
	\begin{equation}
	B^{-1/2} = diag\left(B_{\sO}^{-1/2}, B_{\sBO}^{-1/2},B_{\sT}^{-1/2},B_{\sBT}^{-1/2} \right)
	\end{equation}
	with $B_{\sO} = \comp{B_{1}}{\sO}{\sO} + h I$, $ B_{\sBO} = \comp{B_{1}}{\sBO}{\sBO}$, $B_{\sT} = \comp{B_{2}}{\sT}{\sT}+ h I$, and $B_{\sBT} = \comp{B_{1}}{\sBT}{\sBT}$ and where the term $+h I$ appears due to the interconnection of the two networks.
	Since $B_{1}$ and $B_{2}$ are diagonal, so is $B$, and we get
	\begin{align}
	\hC &= B^{-1/2} A B^{-1/2} = \begin{pmatrix}
	H_{1}(h) & C \\
	C^{T} & H_{2}(h)
	\end{pmatrix},\quad C = \begin{pmatrix}
	\frac{h }{\beta_{1} \beta_{2}} I_{2\times 2}& 0 \\
	0 & 0 
	\end{pmatrix}
	\end{align}
	where $\beta_{i} = \beta_{i}(h) = \sqrt{\alpha_{i} + h}$ [note that since we demanded $\comp{B_{i}}{u_{i}}{u_{i}} = \comp{B_{i}}{v_{i}}{v_{i}} := \alpha_{i}$, we have $B_{\sI} = (\alpha_{i} + h)I$], and with
	\begin{equation}
	H_{i}(h) = \begin{pmatrix}
	\frac{\comp{A}{\sI}{\sI}}{\alpha_{i} + h} & \frac{\comp{A}{\sI}{\sBI} B_{\sBI}^{-1/2}}{\sqrt{\alpha_{i} + h}} \\
	\frac{B_{\sBI}^{-1/2} \comp{A}{\sBI}{\sI}}{\sqrt{\alpha_{i} + h}} & B_{\sBI}^{-1/2} \comp{A}{\sBI}{\sBI} B_{\sBO}^{-1/2}
	\end{pmatrix} := 
	\begin{pmatrix}
	\frac{D_{i}}{\beta^2} & \frac{E_{i}}{\beta} \\
	\frac{E_{i}^T}{\beta} & F_{i}
	\end{pmatrix} .
	\end{equation}
	
	Within $\subsystem{i}$, $u_{i},v_{i}$ are latently mirror symmetric for $h = 0$.
	As can be proven by combining Lemma 11.1 of Ref. \cite{Godsil2017A1StronglyCospectralVertices} with Ref. \cite{Kempton2020LAIA594226CharacterizingCospectralVerticesIsospectral}, latent symmetry is equivalent to the existence of an orthogonal and symmetric matrix $Q_{1}$ which commutes with $H_{1}(h=0)$ and which acts as the permutation $R = \begin{psmallmatrix}
	0 & 1 \\
	1 & 0
	\end{psmallmatrix}$ on $u_{i},v_{i}$ and as an orthogonal transformation $\widetilde{Q}_{1}$ on the remaining sites of $\subsystem{1}$.
	This matrix thus has the form
	\begin{equation}
	Q_{i} = \begin{pmatrix}
	R & 0 \\
	0 & \widetilde{Q}_{i}
	\end{pmatrix}.
	\end{equation}
	We then have
	\begin{equation} \label{eq:commutator}
	\left[Q_{i},H_{i}(h) \right] = \begin{pmatrix}
	\frac{\left[R,D_{i} \right]}{\beta^2} & \frac{R E_{i} - E_{i} \widetilde{Q}_{i}}{\beta} \\
	\frac{\widetilde{Q}_{1} E_{i}^{T} - E_{i}^{T} R}{\beta} & \left[\widetilde{Q}_{i},F_{i} \right]
	\end{pmatrix} \, .
	\end{equation}
	Now, since $\beta(h) > 0$ for all $h > 0$ (since $B$ is diagonal and positive definite, each of its diagonal elements must be greater than zero), it is especially non-zero for $h = 0$.
	Since $\left[Q_{i},H_{i}(h=0) \right] = 0$, we see that $\left[R,D_{i} \right] = 0$, $R E_{i} - E_{i} \widetilde{Q}_{i} = 0$, $\widetilde{Q}_{1} E_{i}^{T} - E_{i}^{T} R = 0$, and also $\left[\widetilde{Q}_{i},F_{i} \right] = 0$.
	Thus, and again since $\beta(h) > 0$ for all $h>0$, we see that the commutator \cref{eq:commutator} vanishes for all $h\le 0$ since it vanishes for $h=0$.
	Thus, $u_{i},v_{i}$ are latently symmetric in $H_{i}(h)$ for any value of $h$.
	
	By concatenating (in the sense of a direct sum) $Q = Q_{1} \oplus Q_{2}$, we can then build a new $Q$-matrix which (i) is orthogonal and symmetric, and which additionally commutes with $\hC$, as is easy to show:
	\begin{equation}
	\left[
	\begin{pmatrix}
	Q_{1} & 0 \\
	0 & Q_{2}
	\end{pmatrix}
	,
	\begin{pmatrix}
	H_{1} & C \\
	C^{T} & H_{2}
	\end{pmatrix} \right] = \begin{pmatrix}
	\left[Q_{1},H_{1} \right] & Q_{1} C - C Q_{2} \\
	Q_{2} C^{T} - C^{T} Q_{1} & \left[Q_{2},H_{2} \right]
	\end{pmatrix} = 0
	\end{equation}
	since $Q_{i} C = C Q_{i} = \frac{h}{\beta_{1} \beta_{2}} R$.
	As a consequence, $u_{i},v_{i}$ are latently mirror symmetric also in $\hC$.
	
	The theorem can then be proven by iteration.
	For example, if we consider the network $\subsystem{1} + \subsystem{2} + \subsystem{3}$, we first apply the theorem to $\subsystem{1} + \subsystem{2}:=\subsystem{1}'$, and then to the combination $\subsystem{1}'+\subsystem{3}$.
	We get that $Q = Q'_{1} \oplus Q_{3}$ commutes with the matrix $\hC$ (now describing $\subsystem{1} + \subsystem{2} + \subsystem{3}$).
	Now, since $Q'_{1} = Q_{1} \oplus Q_{2}$, we have $Q = Q_{1} \oplus Q_{2} \oplus Q_{3}$.
	$Q$ thus acts as the permutation $R$ individually on each $S_{i}$, which is equivalent to the statement that each $S_{i}$ is latently mirror symmetric in $\hC$.
\end{proof}

\section{Domain-wise parity} \label{sec:domainWiseParity}

In the last part of the main manuscript, we demonstrated how a setup with no global reflection symmetry can be built such that the low-frequency eigenmodes have domain-wise parity.
In the following, we will show the main logic behind this construction principle.

\begin{figure}[htb] 
	\centering
	\includegraphics[width=0.4\textwidth]{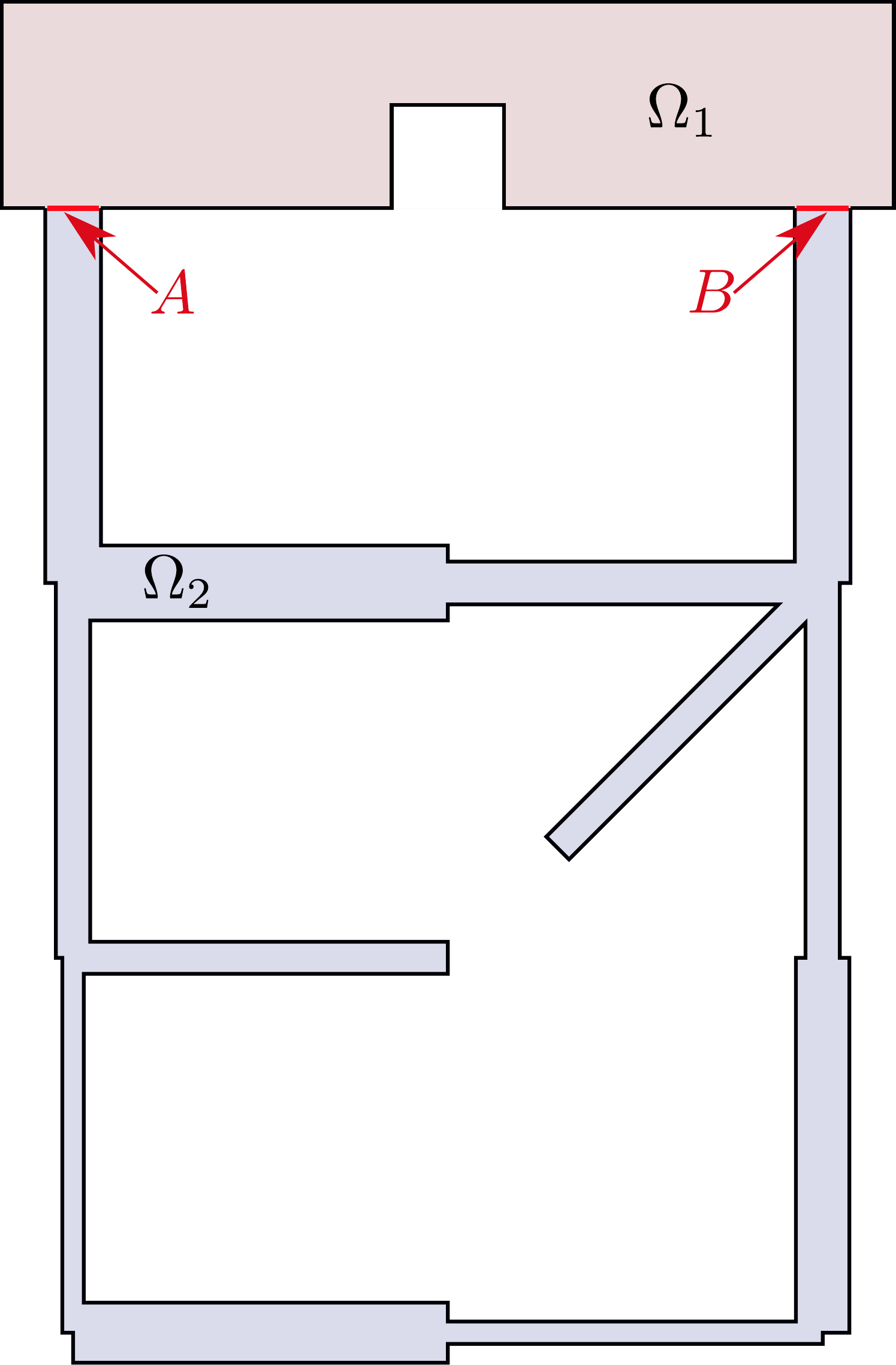}
	\caption{The division of the total setup into two different domains $\Omega_{1}$ and $\Omega_{2}$.
	}
	\label{fig:divisionOfDomains}
\end{figure}

To start, we assume that the lower waveguide network in \cref{fig:divisionOfDomains} has been extended by two identical waveguides, each of them connected to one of the two latently symmetric points.
As can easily be shown through \cref{thm:interconnection}, the two upper endpoints of these waveguides are latently symmetric as well.
In the following $\Omega_{2}$ will denote the lower waveguide network (extended by the two connecting waveguides), and $\Omega_{1}$ will denote the upper cavity.

Before the two subsystems $\Omega_{1}$ and $\Omega_{2}$ are coupled to each other, the 3D-Helmholtz equation in each of these two domains reads
\begin{align}
&\left(\Delta + \alpha^2\right)\phi = 0\; \text{within the interior of $\Omega_{1}$ and $\partial_{\mathbf{n}}\, \phi = 0$ on $\partial\Omega_{1}$} \\
&\left(\Delta + \beta^2\right)\psi = 0\; \text{within the interior of $\Omega_{2}$ and $\partial_{\mathbf{n}}\, \psi = 0$ on $\partial\Omega_{2}$} \, .
\end{align}
That is, the solution has to fulfill Neumann hard boundary (wall) conditions on the two segments $A$ and $B$.

After coupling the two subsystems to each other, the 3D-Helmholtz equation for the full setup reads
\begin{equation}
\left(\Delta + k^2\right)p = 0\; \text{within the interior of $\left( \Omega_{1} \cup \Omega_{2}\right)$ and $\partial_{\mathbf{n}}\, p = 0$ on $\partial\left( \Omega_{1} \cup \Omega_{2}\right)$} \, .
\end{equation}
where $\partial\left( \Omega_{1} \cup \Omega_{2}\right)$ no longer includes the two segments $A$ and $B$.

We are interested in the behavior of $p$ at the intersection of $\Omega_{1}$ and $\Omega_{2}$, that is, on $A$ and $B$.
To this end, we expand $p$ on both $A$ and $B$ in terms of the orthonormal bases of $\{\phi_{n} \}$ and the $\{\psi_{m} \}$.
Before doing so, we remind the reader that these fulfill
\begin{align}
&\left(\Delta + \alpha^2\right)\phi = 0\; \text{within the interior of $\Omega_{1}$ and $\partial_{\mathbf{n}}\, \phi = 0$ on $\partial\Omega_{1}$} \\
&\left(\Delta + \beta^2\right)\psi = 0\; \text{within the interior of $\Omega_{2}$ and $\partial_{\mathbf{n}}\, \psi = 0$ on $\partial\Omega_{2}$} \, .
\end{align}
and
\begin{align}
	\int_{\Omega_{1}} \phi_{n}^{*} \phi_{n'}\, dV &= \delta_{n,n'} \\
	\int_{\Omega_{2}} \psi_{m}^{*} \psi_{m'}\, dV &= \delta_{m,m'}
\end{align}
with $\delta_{n,n'}$ being the Kronecker delta.
Using $\{\phi_{n} \}$ and $\{\psi_{m} \}$ to expand $p$ on $A,B$, we obtain
\begin{align}
p &= \sum_{n} a_{n} \phi_{n} \; \text{within $\Omega_{1}$}\\
p &= \sum_{m} b_{m} \psi_{m} \; \text{within $\Omega_{2}$}
\end{align}
with $a_{n} = \int_{\Omega_{1}} p\, \phi^{*}_{n}$, $b_{m} = \int_{\Omega_{2}} p\, \psi^{*}_{m}$, and $	\int_{\Omega_{1}} \phi^{*}_{k} \phi_{l} = \int_{\Omega_{2}} \psi^{*}_{k} \psi_{l} = \delta_{k l}$ and where the star $*$ denotes complex conjugation.

To obtain a relation for $a_{n}$, we compute the following integral
\begin{align}
	\int_{\Omega_{1}} \phi_{n}^{*} \Delta p\, dV &= \oint_{\partial\Omega_{1}} \left(\phi_{n}^{*} \partial_{\mathbf{n}} p - p \partial_{\mathbf{n}} \phi_{n}^{*} \right) + \int_{\Omega_{1}} p\, \Delta \phi_{n}^{*}\, dV \\
	&= \oint_{A+B} \phi_{n} \partial_{\mathbf{n}} p - \alpha_{n}^2 \int_{\Omega_{1}} p \phi_{n}^{*}\, dV = \oint_{A+B} \phi_{n} \partial_{\mathbf{n}} p - a_{n} \alpha_{n}^2 \\
	&= k^2 \int_{\Omega_{1}} \phi_{n}^{*} p \, dV = k^2 a_{n} 
	\,.
\end{align}
For small waveguide cross sections, that is, in the monomode approximation, the contour integral over the two surfaces $A,B$ (which then shrink to the single points $A$, $B$) becomes
\begin{equation}
	S_{A} \partial_{n}^{(\Omega_{1})} p(A) \phi_{n}(A) + S_{B} \partial_{n}^{(\Omega_{1})} p(B) \phi_{n}(B)
\end{equation}
where $S_{A}$, $S_{B}$ denote the waveguide cross section at the two points $A,B$ and where we have used the short-hand notation $\partial_{n}^{(\Omega_{1})} p(\mathbf{x})$ to denote $\partial_{\mathbf{n}} p(\mathbf{r})\vert_{\mathbf{r} = \mathbf{x}}$ with $\mathbf{n}$ denoting the normal to the boundary $\partial \Omega_{1}$.

Assuming that $k^2 \ne \alpha_{n}^2$, the above gives us
\begin{equation}
	a_{n} = S \frac{ \partial_{n}^{(\Omega_{1})} p(A) \phi_{n}(A) + \partial_{n}^{(\Omega_{1})} p(B) \phi_{n}(B)}{\alpha_{n}^2 - k^2}
\end{equation}
where we have used the fact that the two cross sections are identical by symmetry, that is, $S_{A} = S_{B} \equiv S$.
In a completely analogous manner, we obtain
\begin{equation}
b_{m} = S \frac{ \partial_{n}^{(\Omega_{2})} p(A) \psi_{m}(A) + \partial_{n}^{(\Omega_{2})} p(B) \psi_{m}(B)}{\beta_{m}^2 - k^2} \, .
\end{equation}

Inserting the expression for $a_{n}$ into the eigenmode expansion of $p$ within $\Omega_{1}$, we yield (with $\gamma$ being either $A$ or $B$)
\begin{align}
	p(\gamma) = \sum_{n} a_{n} \phi_{n}(\gamma) &= S \frac{ \partial_{n}^{(\Omega_{1})} p(A) \phi_{n}(A) + \partial_{n}^{(\Omega_{1})} p(B) \phi_{n}(B)}{\alpha_{n}^2 - k^2} \phi_{n}(\gamma) \, .
\end{align}
Defining $G_{n} \equiv \frac{S}{\alpha_{n}^2 - k^2}$, we can write these two equations (one for $\gamma = A$ and one for $\gamma = B$) as the impedance relation
\begin{equation}
	\begin{pmatrix}
	p(A)\\
	p(B)
	\end{pmatrix} = 
\underbracket{	\begin{pmatrix}
	\sum_{n} G_{n} \phi_{n}^2(A) & \sum_{n} G_{n} \phi_{n}(A) \phi_{n}(B) \\
	\sum_{n} G_{n} \phi_{n}(A) \phi_{n}(B) & \sum_{n} G_{n} \phi_{n}^2(B)
	\end{pmatrix}}_{\equiv M_{C}}
	\begin{pmatrix}
	\partial_{n}^{(\Omega_{1})} p(A) \\
	\partial_{n}^{(\Omega_{1})} p(B)
	\end{pmatrix} \, .
\end{equation}

Similarly, by performing the above steps again for $\Omega_{2}$, we get
\begin{align}
\begin{pmatrix}
p(A)\\
p(B)
\end{pmatrix} & = 
\underbracket{\begin{pmatrix}
	\sum_{m} \xi_{m} \psi_{m}^2(A) & \sum_{m} \xi_{m} \psi_{m}(A) \psi_{m}(B) \\
	\sum_{m} \xi_{m} \psi_{m}(A) \psi_{m}(B) & \sum_{m} \xi_{m} \psi_{m}^2(B)
	\end{pmatrix}}_{\equiv Z_{\Omega_2}}
\begin{pmatrix}
\partial_{n}^{(\Omega_{2})} p(A) \\
\partial_{n}^{(\Omega_{2})} p(B)
\end{pmatrix} \\
&=
- Z_{\Omega_2} 	\begin{pmatrix}
\partial_{n}^{(\Omega_{1})} p(A) \\
\partial_{n}^{(\Omega_{1})} p(B)
\end{pmatrix} \label{eq:impedanceRelation}
\end{align}
with $\xi_{m} \equiv \frac{S}{\beta_{m}^2 - k^2}$ and where we have used the fact that
\begin{equation} \label{eq:derivativeRelation}
	\begin{pmatrix}
\partial_{n}^{(\Omega_{2})} p(A) \\
\partial_{n}^{(\Omega_{2})} p(B)
\end{pmatrix}
= - \begin{pmatrix}
	\partial_{n}^{(\Omega_{1})} p(A) \\
	\partial_{n}^{(\Omega_{1})} p(B)
	\end{pmatrix} \, .
\end{equation}

Due to (latent) symmetry, we have $\phi_{n}(A) = \pm \phi_{n}(B)$, and also $\psi_{m}(A) = \pm \psi_{m}(B)$, and we see that both matrices $Z_{\Omega_1}$ and $Z_{\Omega_2}$ are bi-symmetric, that is, they have the structure $\begin{psmallmatrix}
a & b \\
b & a
\end{psmallmatrix}$.
Moreover, one can show that both of these matrices must be invertible.
We thus get
\begin{equation}
\begin{pmatrix}
p(A) \\
p(B)
\end{pmatrix}
=
- Z_{\Omega_1}^{-1} Z_{\Omega_2} \begin{pmatrix}
p(A) \\
p(B)
\end{pmatrix}
\end{equation}
so that $\begin{psmallmatrix}
p(A) \\
p(B)
\end{psmallmatrix}$ is an eigenvector of the bi-symmetric, $k$-dependent matrix $Z_{\Omega_1}^{-1} Z_{\Omega_2}$.
Now, since the eigenvectors of bi-symmetric matrices are $\begin{psmallmatrix}
1 \\
\pm 1
\end{psmallmatrix}$, we see that $p(A) = \pm p(B)$.
From there, it is trivial to show that $p$ also has parity in $\Omega_{1}$ as a whole.

\subsection{Impedance plots for Figure 2 (c1)} \label{sec:EigenvaluePlots}

\begin{figure}[htb] 
	\centering
	\includegraphics[width=0.25\textwidth]{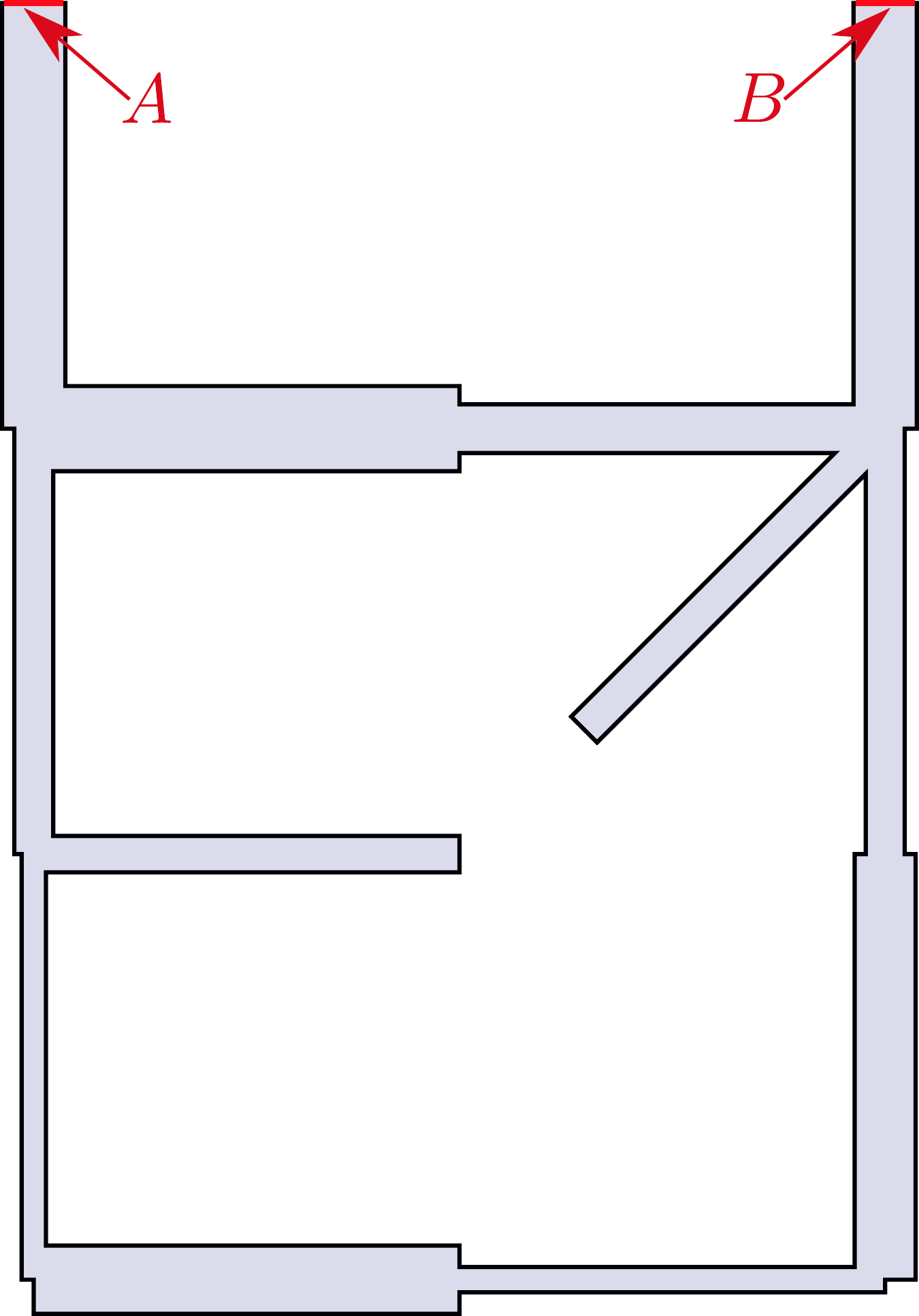}
	\caption{The setup (with open ports $A$, $B$) for which the impedance is calculated.
	}
	\label{fig:impedanceDomains}
\end{figure}

In this subsection, we investigate the impedance of the system depicted in \cref{fig:impedanceDomains}.
Geometrically, this setup corresponds to the one shown in Fig. 2 (c1) of the manuscript, though with the upper coupled cavity removed.

In order to compute the impedance, we open the ports $A,B$ and investigate plane-wave scattering of this setup.
If we send in a plane-wave with amplitude $a_{A}$ and wavenumber $k$ into port $A$, a portion with amplitude $b_{A}$ will reflected, while a portion with amplitude $c_{A,B}$ will leave the system at port $B$. In complete analogy, if we send in a plane-wave with amplitude $a_{B}$ and wavenumber $k$ into port $B$, a portion with amplitude $b_{B}$ will reflected, while a portion with amplitude $c_{B,A}$ will leave the system at port $A$.
We thus obtain the following scattering matrix
\begin{equation}
	S = \begin{pmatrix}
		r_{A} & t_{AB} \\
		t_{AB} & r_{B}
	\end{pmatrix}
\end{equation}
with $r_{A} = b_{A}/a_{A}$, $r_{B} = b_{B}/a_{B}$, and $t_{AB} = c_{A,B}/a_{A} = c_{B,A}/a_{B}$.

The impedance matrix, connecting the pressure $\mathbf{p} = (p_{A},p_{B})^{T}$ and its normal derivative $\mathbf{p}' = (p'_{A},p_{B}')^T$ at the points $A,B$ via $\mathbf{p} = Z \mathbf{p}'$, reads (using that $p_{n} = a_{n} + b_{n}$ and $p'_{n} = i k (a_{n} - b_{n})$)
\begin{equation}
	Z = \frac{1}{i k} (I + S)(I-S)^{-1}= \begin{pmatrix}
		Z_{AA} & Z_{AB} \\ 
		Z_{BA} & Z_{BB} 
	\end{pmatrix} = \beta
	\begin{pmatrix}
		\alpha - \delta & 2 t_{AB}\\
		2 t_{AB} & \alpha + \delta 
	\end{pmatrix}	
\end{equation}
with $I$ denoting the two-by-two identity matrix, $\delta = r_{A}-r_{B}$, $\alpha = |S| -  1$, and $\beta = \left(i k (Tr(S) - |S| - 1)\right)^{-1}$, and where $|S|, Tr(S)$ denote the determinant and trace of $S$, respectively.

In the following, we plot a comparison of $Z_{AA},Z_{BB}$ for the setup depicted in \cref{fig:impedanceDomains} for different ratios $w_{max}/L$.

\begin{figure}[htb] 
	\centering
	\includegraphics[width=0.75\textwidth]{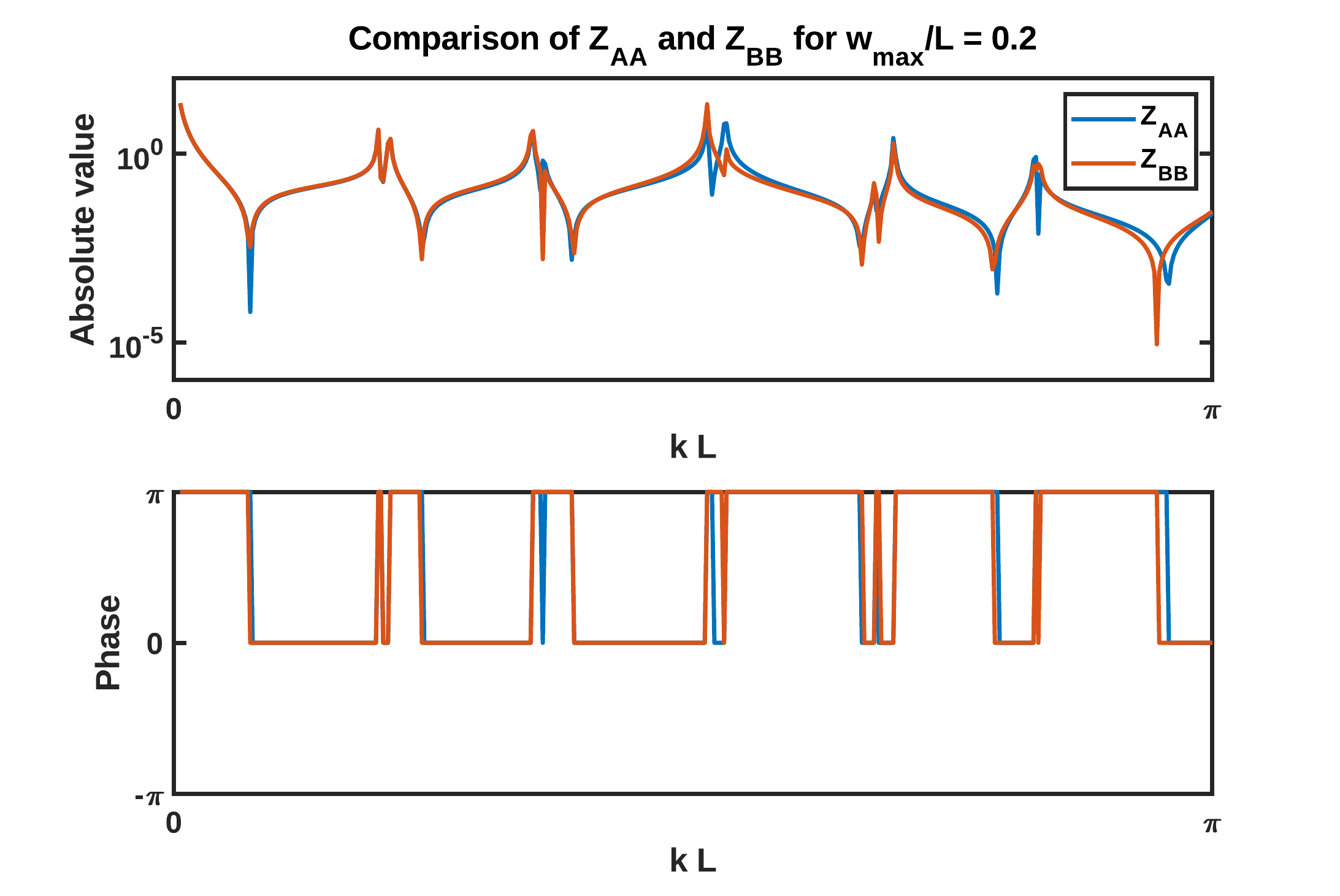}
\end{figure}
\begin{figure}[htb] 
	\centering
	\includegraphics[width=0.75\textwidth]{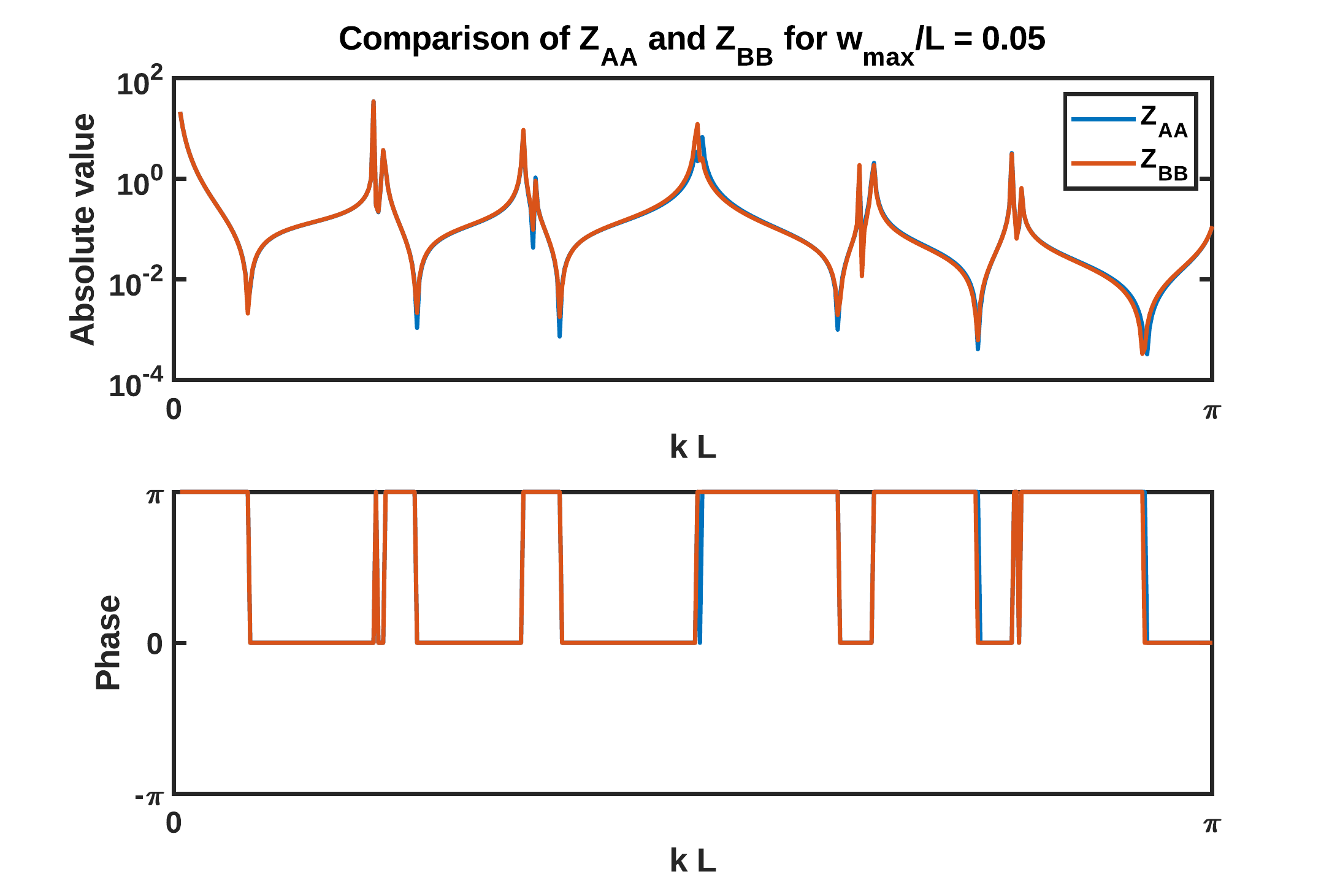}
\end{figure}

\begin{figure}[htb] 
	\centering
	\includegraphics[width=0.75\textwidth]{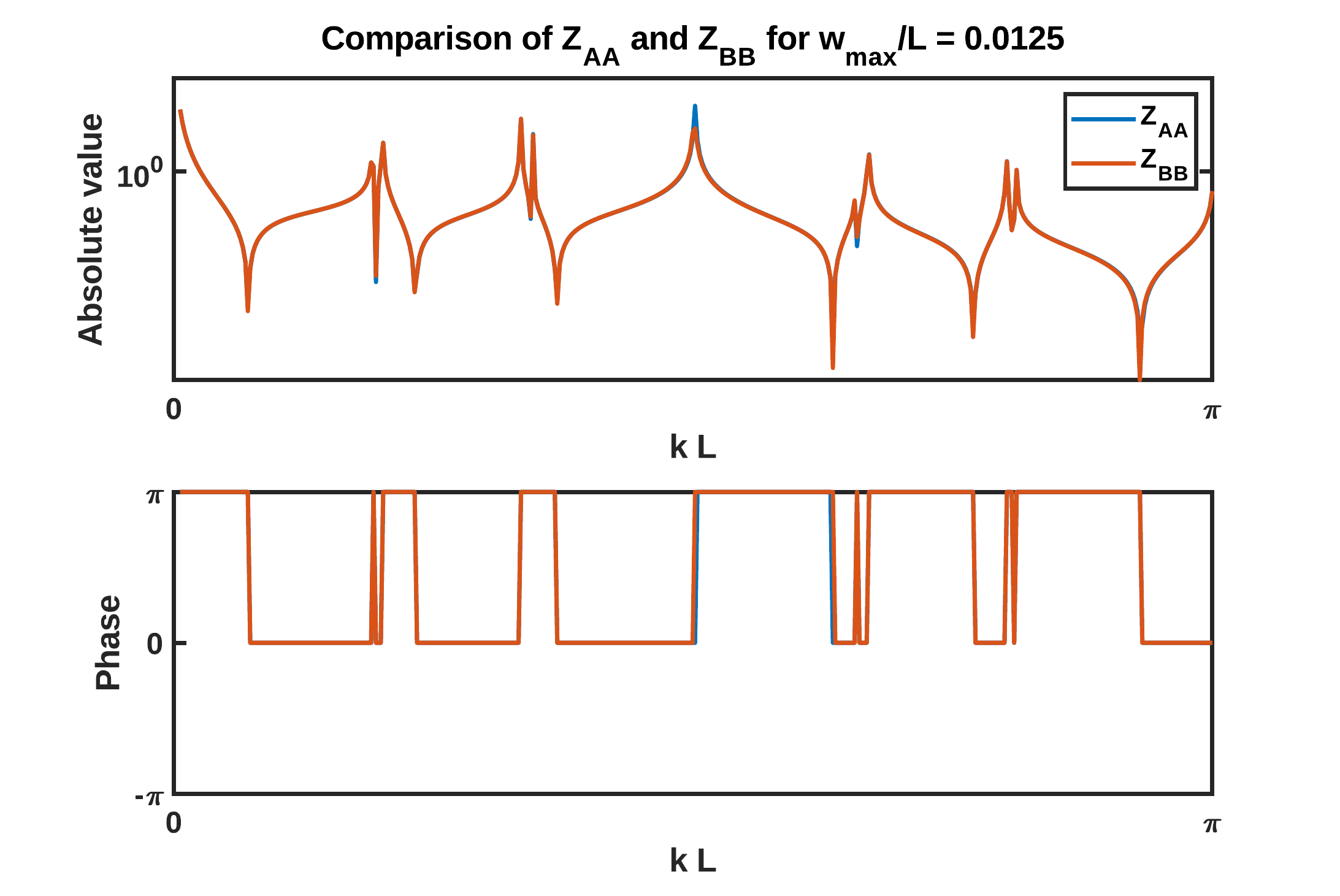}
\end{figure}

\clearpage

%


%